\documentclass[11pt]{article}

\usepackage{verbatim}
\usepackage{amsmath}
\usepackage{amsfonts}
\usepackage{amssymb}
\usepackage{bbm}
\usepackage{latexsym}
\usepackage{graphics}
\usepackage{lscape} 
\usepackage{epsfig}
\usepackage{color}

\usepackage{cite}

\renewcommand{\d}{\mathrm{d}}

\newcommand{\captn}[1]{\vspace{-3ex}\caption{\small #1}}

\DeclareMathSymbol{\mg}{\mathrel}{symbols}{"1D}

\renewcommand{\gg}{\gamma}
\newcommand{\gd}{\delta}
\renewcommand{\ge}{\epsilon}

\newcommand{\gf}{\phi}

\newcommand{\gl}{\lambda}

\newcommand{\gth}{\theta}

\newcommand{\go}{\omega}

\newcommand{\gp}{\pi}

\newcommand{\gTh}{\Theta}

\newcommand{\cA}{{\cal A}}

\newcommand{\cF}{{\cal F}}

\newcommand{\cK}{{\cal K}}

\newcommand{\cR}{{\cal R}}

\newcommand{\cV}{{\cal V}}

\newcommand{\tZ}{{\tilde Z}}
\newcommand{\tr}{\text{tr}}

\newcommand{\ra}{\rightarrow}

\newcommand{\inv}{^{-1}}
\newcommand{\dsp}{\displaystyle}

\newcommand{\labl}[1]{\label{#1}}
\newcommand{\Kh}{K\"{a}hler}
\newcommand{\beq}{\begin{equation}}
\newcommand{\eeq}{\end{equation}}
\newcommand{\barr}{\begin{array}}
\newcommand{\earr}{\end{array}}
\newcommand{\equ}[1]{\begin{gather} #1 \end{gather}}

\newcommand{\enums}[1]{\begin{enumerate} #1 \end{enumerate}}

\newcommand{\arry}[2]{\begin{array}{#1} #2 \end{array}}

\newcommand{\pmtrx}[1]{\begin{pmatrix} #1 \end{pmatrix}}

\newcounter{oldcounter}

\newcommand{\fA}{\mathfrak{ A}}

\newcommand{\bee}{{\bar e}}

\newcommand{\bz}{{\bar z}}

\newcommand{\bZ}{{\bar Z}}
\newcommand{\bge}{{\bar\epsilon}}

\newcommand{\Intr}{\mathbb{Z}}
\newcommand{\Cplx}{\mathbb{C}}

\newcommand{\CP}{\mathbb{CP}}

\newcommand{\qand}{\quad \text{and} \quad}

\newcommand{\Res}{\text{Res}}

\newcommand{\ba}[2]{\[\begin{array}{#2}\label{#1}}
\newcommand{\ea}{\end{array}\]}
\newcommand{\be}{\begin{equation}}
\newcommand{\ee}{\end{equation}}
\newcommand{\bea}{\begin{eqnarray}}
\newcommand{\eea}{\end{eqnarray}}

\newcommand{\E}[1]{\mathrm{E_{#1}}}
\newcommand{\U}[1]{\mathrm{U(#1)}}
\newcommand{\SU}[1]{\mathrm{SU(#1)}}
\newcommand{\SO}[1]{\mathrm{SO(#1)}}

\newcommand{\rep}[1]{\mathbf{#1}}
\newcommand{\crep}[1]{\overline{\rep{#1}}}

\def\intersectionsCtwoZthree{
\begin{table} 
\begin{center} 
\begin{tabular}{| c || c | c | c | c |}
\hline &&&& \\ [-2ex]
divisor & $D_1$ & $D_2$ & $E_1$ & $E_2$ 
\\[1ex]\hline\hline &&&& \\ [-2ex]
$E_1$ & $0$ & $1$ & $-2$ & $1$ 
\\[1ex]\hline &&&& \\ [-2ex]
$E_2$ & $1$ & $0$ & $1$ & $-2$ 
\\[1ex]\hline\hline &&&& \\ [-2ex]
$D_1$ & $-\frac 23$ & $-\frac 13$ & $0$ & $1$ 
\\[1ex]\hline &&&& \\ [-2ex]
$D_2$ & $-\frac 13$ & $-\frac 23$ & $1$ & $0$ 
\\[1ex]\hline
\end{tabular} 
\end{center}
\captn{\label{tb:intersectionsC2Z3}
The upper half of the table gives intersection numbers of the compact
curves, $E_1$ and $E_2$, with all divisors of the resolution
$\Res(\Cplx^2/\Intr_3)\,$.  The bottom half of the table gives the
values of the integrals over the product of the $(1,1)$--forms
corresponding to the divisors, which are not necessarily integral. 
}
\end{table}
}

\def\modelsCtwoZthree{
\begin{table} 
\begin{center} 
\begin{tabular}{| c || c | c  |}
\hline && \\ [-2ex]
orbifold & blowup & blowup 
\\ [-1ex] & & \\[-2ex] 
shift $3\,v$ & vector $V_1$ & vector $V_2$ 
\\[1ex]\hline\hline && \\ [-2ex]
$(1^2, 0^{14})$ & $(2^2, 0^{14})$ & $-(2^2, 0^{14})$
\\[1ex] && \\ [-2ex]
$$ & $(2,1, 0^{14})$ & $(1, -1, 0^{14})$
\\[1ex]\hline && \\ [-2ex]
$(2, 1^4, 0^{11})$ & $(2,1^4, 0^{11})$ & $-(2, 1^4, 0^{11})$
\\[1ex]\hline && \\ [-2ex]
$(1^8, 0^{8})$ & $(1^8, 0^{8})$ & $-(1^8, 0^{8})$
\\[1ex]\hline && \\ [-2ex]
$(1^{14}, 0^{2})$ & $\frac 12(1^{14}, 3^{2})$ & $-\frac12(1^{14}, 3^{2})$
\\[1ex]\hline\hline && \\ [-2ex]
$(2, 1^{10},0^5)$ & $(2,1^{10},0^5)$ & $-(2,1^{10},0^5)$ 
\\[1ex] && \\ [-2ex]
& $\frac 12 (-3, 1^{10}, 1^5)$ & $(1,0^{10},-1^5)$
\\[1ex]\hline
\end{tabular} 
\end{center}
\captn{\label{tb:modelsC2Z3}
This table compares the $\Cplx^2/\Intr_3$ orbifold gauge shift vector
$v$ with the blowup vectors, $V_1$ and $V_2\,$, that topologically
characterize gauge background  of the resolution
$\Res(\Cplx^2/\Intr_3)\,$. The blowup vectors under the double line do
not satisfy all possible conditions simultaneously.  The upper proposal
gives a non--vanishing Bianchi, while the vectors of the bottom one
cannot be identified with the orbifold shift. 
}
\end{table}
}

\def\intersectionsCthreeZfour{
\begin{table} 
\begin{center} 
\begin{tabular}{| c || c | c | c | c | c |}
\hline &&&&& \\ [-2ex]
  & $D_1$ & $D_2$ & $D_3$ & $E_1$ & $E_2$ 
\\[1ex]\hline\hline &&&&& \\ [-2ex]
$D_1 E_1$ & $0$ & $0$ & $1$ & $-2$ & $1$ 
\\[1ex]\hline &&&&& \\ [-2ex]
$D_2 E_1$ & $0$ & $0$ & $1$ & $-2$ & $1$ 
\\[1ex]\hline &&&&& \\ [-2ex]
$D_3 E_1$ & $1$ & $1$ & $2$ & $-4$ & $0$ 
\\[1ex]\hline &&&&& \\ [-2ex]
$E_1 E_2$ & $1$ & $1$ & $0$ & $0$ & $-2$ 
\\[1ex]\hline\hline &&&&& \\ [-2ex]
$D_3 E_2$ & $-\frac 12$ & $-\frac 12$ & $0$ & $0$ & $1$ 
\\[1ex]\hline
\end{tabular} 
\end{center}
\captn{\label{tb:intersectionsC3Z4}
The first part of the table gives all possible intersection numbers of
the compact curves with all divisors of the resolution
$\Res(\Cplx^3/\Intr_4)\,$. As the curve $D_3 E_2$ is excluded, 
the final row of this table can only be interpret as giving 
(fractional) values of the integrals of the corresponding forms. 
}
\end{table}
}

\def\modelsCthreeZfour{
\begin{table}
\begin{center}
\begin{tabular}{c c}
\begin{tabular}[t]{| c | c  c | l |}
\hline &&& \\ [-2ex]
orbifold & blowup & blowup &
\\ [-1ex] &&& \\[-2ex]
shift $4\,v$ & vector $V_2$ & vector $V_1$  & Nr. 
\\[1ex]\hline\hline &&& \\ [-2ex]
$(0^{13},1^2,2)$ 
& $(0^{13},1^2,2)$ 
& $(0^{13},1^2,$-$2)$ 
& 1a
\\[.7ex]
& $(0^{13},1^2,2)$
& $(0^{12},2,$-$1^2,0)$
& 1b
\\[.7ex]
& $(0^{13},1^2,2)$
& $(0^{11},2,1,0^2,$-$1)$
& 1c  
\\[1ex]\hline &&& \\ [-2ex]
$(0^{11},1^2,2^3)$ 
& $(0^{13},1^2,2)$ 
& $(0^{10},1^4,$-$1^2)$
& 2a 
\\[.7ex]
&  $(0^{13},1^2,2)$
&  $(0^{11},1^2,$-$2,0^2)$
& 2b 
\\[1ex]\hline &&& \\ [-2ex]
$(0^{9},1^2,2^5)$ 
& $(0^{13},1^2,2)$
& $(0^{8},1^5,0^2,$-$1)$
& 3a 
\\[.7ex]
&  $(0^{13},1^2,2)$
&  $(0^{9},1^4,$-$1^2,0)$
& 3b  
\\[1ex]\hline &&& \\ [-2ex]
$(0^{7},1^2,2^7)$ & $-$ & $-$ & 4 
\\[1ex]\hline &&& \\ [-2ex]
$(0^{10},1^6)$ 
& $(0^{10},1^6)$ 
& $(0^{10},1^2,$-$1^4)$ 
& 5a
\\[.7ex]
&  $(0^{10},1^6)$
&  $(0^{13},1,$-$1,$-$2)$
& 5b  
\\[1ex]\hline &&& \\ [-2ex]
$(0^{10},1^5,3)$ 
& $(0^{10},1^6)$
& $(0^{9},2,$-$1^2,0^4)$
& 6 
\\[1ex]\hline &&& \\ [-2ex]
$(0^{8},1^6,2^2)$ 
& $(0^{10},1^6)$
& $(0^{8},1^3,$-$1^3,0^2)$
& 7a 
\\[.7ex]
&  $(0^{10},1^6)$
&  $(0^{8},1^2,$-$2,0^5)$
& 7b  
\\[1ex]\hline &&& \\ [-2ex]
$(0^{6},1^6,2^4)$ 
& $(0^{10},1^6)$
& $(0^{6},1^4,$-$1^2,0^4)$
& 8 
\\[1ex]\hline
\end{tabular} \hspace{-14pt} &
\begin{tabular}[t]{| c | c c | c |}
\hline &&& \\ [-2ex]
orbifold & blowup & blowup &
\\ [-1ex] &&& \\[-2ex]
shift $4\,v$ & vector $V_2$ & vector $V_1$  & Nr. 
\\[1ex]\hline\hline &&& \\ [-2ex]
$(0^{5},1^{10},2)$ 
&  $(0^{10},1^6)$
& $\frac{1}{2}($-$3,1^{10},$-$1^5)$
& 9 
\\[1ex]\hline &&& \\ [-2ex]
$(0^{3},1^{10},2^3)$ 
& $(0^{10},1^6)$
& $\frac{1}{2}(1^{12},$-$1^3,$-$3)$
& 10 
\\[1ex]\hline &&& \\ [-2ex]
$(1^{14},2^2)$ 
& $(0^{13},$-$2,1^2)$
& $\frac{1}{2}(1^{15},$-$3)$
& 11 
\\[1ex]\hline &&& \\ [-2ex]
$(1^{13},$-$1,2^2)$ 
& $(0^{13},1^2,2)$
& $\frac{1}{2}(1^{15},$-$3)$
& 12a 
\\[.7ex]
& $(0^{13},1^2,2)$
& -$\frac{1}{2}($-$3,1^{15})$
& 12b 
\\[1ex]\hline &&& \\ [-2ex]
$\frac{1}{2}(1^{3},3^{12},$-$3)$ 
& $\frac{1}{2}($-$3,1^{15})$
& $$-$(0^{13},1^2,2)$
& 13a 
\\[.7ex]
& $\frac{1}{2}(1^{15},$-$3)$
& $(0^{13},1^2,2)$
& 13b 
\\[.7ex]
& $\frac{1}{2}(1^{15},$-$3)$
& $\frac{1}{2}(1^3,$-$1^{11},3,1)$
& 13c 
\\[1ex]\hline &&& \\ [-2ex]
$\frac{1}{2}(1^{7},3^{8},$-$3)$ 
& $\frac{1}{2}(1^{15},$-$3)$
& $($-$1^5,1,0^{10})$
& 14a 
\\[.7ex]
& $\frac{1}{2}(1^{15},$-$3)$
& $\frac{1}{2}(1^{6},$-$1^8,$-$3,1)$
& 14b 
\\[1ex]\hline &&& \\ [-2ex]
$\frac{1}{2}(1^{11},3^{4},$-$3)$ 
& $\frac{1}{2}(1^{15},$-$3)$
& $(0^{10},1^3,$-$1^3)$
& 15 
\\[1ex]\hline &&& \\ [-2ex]
$\frac{1}{2}(1^{15},$-$3)$ 
& $\frac{1}{2}(1^{15},$-$3)$
& $(0^{13},$-$2,1^2)$
& 16a 
\\[.7ex]
& $\frac{1}{2}(1^{15},$-$3)$
& $\frac{1}{2}($-$1^{14},3,$-$1)$
& 16b 
\\[1ex]\hline 
\end{tabular}
\end{tabular}
\end{center}
\captn{\label{tb:modelsC3Z4}
This table compares the $\Cplx^3/\Intr_4$ orbifold gauge shift vector $v\,$, 
with the blowup vectors $V_1$ and $V_2\,$,
that characterize the line bundle gauge background on the resolution. 
We provide a complete classification of $U(1)$ fluxes
compatible with the resolution of a $\Cplx^3/\Intr_4$ singularity, i.e.
fulfilling the orbifold matiching~\eqref{identifyZ4shiftbis} and the
Bianchi identities~\eqref{NESbianchiC3Z4}  and~\eqref{EXTRAbianchiC3Z4}.
}
\end{table}
}

\newcounter{savetable}
\newcommand{\subtablenr}{\setcounter{savetable}{\value{table}}
\stepcounter{savetable}\setcounter{table}{0}
\renewcommand{\thetable}{\mbox{\arabic{savetable}.\alph{table}}}
}
\newcommand{\resettablenr}{\setcounter{table}{\value{savetable}}
\renewcommand{\thetable}{\mbox{\arabic{table}}}
}

\def\spectraCthreeZfour{
\subtablenr
\thispagestyle{empty} 
\begin{landscape}
\begin{table}
\begin{center}
\begin{tabular}{| c || c | c | c | c |}
\hline &&&& \\ [-2ex]
Nr. & 4D gauge group & $\frac 18 \times$ ``untwisted''  & 
$\frac 14 \times$ ``2nd twisted'' & ``1st twisted'' 
\\[1ex]\hline\hline &&&& \\ [-2ex]
1a & $\SO{26}\times\U{2}\times\U{1}$ & 
$(\rep{26},\rep{2}) + 2(\rep{1},\rep{2})$ & 
$(\rep{26},\rep{1}) +  2 (\rep{1},\rep{2}) + (\rep{1},\rep{1}) $ & 
$ (\rep{26},\rep{1})+ 2 (\rep{1},\rep{2}) + 3  (\rep{1},\rep{1})$ 
\\[1ex]\hline &&&& \\ [-2ex]
1b & $\SO{24} \times \U{2} \times \U{1}^2$ & 
$ (\rep{24},\rep{2}) + 4  (\rep{1},\rep{2})$ &
$ (\rep{24},\rep{1}) + 2 (\rep{1},\rep{2}) + 3 (\rep{1},\rep{1}) $ &  
$ (\rep{24},\rep{1}) +2 (\rep{1},\rep{2})+ 5  (\rep{1},\rep{1})$  
\\[1ex]\hline &&&& \\ [-2ex]
1c & $\SO{22} \times \U{2} \times \U{1}^3$ & 
$ (\rep{22},\rep{2}) + 6  (\rep{1},\rep{2}) $ &
$ (\rep{22},\rep{1}) +  2(\rep{1},\rep{2}) + 5 (\rep{1},\rep{1}) $ &  
$ (\rep{22},\rep{1}) + 2 (\rep{1},\rep{2})+5  (\rep{1},\rep{1})$  
\\[1ex]\hline &&&& \\ [-2ex]
2a & $\SO{20} \times \U{3} \times \U{1}^3$ &
$\arry{c}{2 (\rep{20},\rep{1}) + 2 (\rep{1},\rep{3})\\
2 (\rep{1},\crep{3}) + 4 (\rep{1},\rep{1})}$ & 
$\arry{c}{(\rep{20},\rep{1}) +  (\rep{1},\rep{3}) + \\ 
(\rep{1},\crep{3})  + 3 (\rep{1},\rep{1})}$ & 
$2 (\rep{1},\rep{3}) +2 (\rep{1},\crep{3}) + 2 (\rep{1},\rep{1})$
\\[1ex]\hline &&&& \\ [-2ex]
2b & $\SO{22} \times \U{2} \times \U{1}^3$ &
$2 (\rep{22},\rep{1}) + 4 (\rep{1},\rep{2}) + 4 (\rep{1},\rep{1})$ & 
$(\rep{22},\rep{1}) + 2 (\rep{1},\rep{2}) + 3 (\rep{1},\rep{1})$ & 
$2 (\rep{1},\rep{2}) + 7 (\rep{1},\rep{1})$
\\[1ex]\hline &&&& \\ [-2ex]
3a & $\SO{16}\times\U{2}\times \U{5}\times\U{1}$ & 
$\arry{c}{(\rep{16},\rep{2},\rep{1}) + (\rep{1},\rep{2},\rep{5})\\ +
(\rep{1},\rep{2},\crep{5}) + 2(\rep{1},\rep{2},\rep{1})}$ & 
$ \arry{c}{(\rep{16},\rep{1},\rep{1}) + (\rep{1},\rep{1},\rep{5})\\ +
(\rep{1},\rep{1},\crep{5})  +
(\rep{1},\rep{1},\rep{1})}$ & 
$  (\rep{1},\rep{1},\rep{10}) + (\rep{1},\rep{1},\crep{5}) $
\\[1ex]\hline &&&& \\ [-2ex]
3b & $\SO{18}\times\U{2}\times \U{4}\times\U{1}$ & 
$\arry{c}{(\rep{18},\rep{2},\rep{1}) + (\rep{1},\rep{2},\rep{4}) \\+
(\rep{1},\rep{2},\crep{4})+2 (\rep{1},\rep{2},\rep{1})}$ & 
$ \arry{c}{(\rep{18},\rep{1},\rep{1}) + (\rep{1},\rep{1},\rep{4}) +
(\rep{1},\rep{1},\crep{4}) \\ + (\rep{1},\rep{2},\rep{1}) +
(\rep{1},\rep{1},\rep{1})}$ & 
$  (\rep{1},\rep{1},\rep{1}) + (\rep{1},\rep{6},\rep{1}) $
\\[1ex]\hline &&&& \\ [-2ex]
5a & $\SO{20}\times\U{4}\times\U{2}$ & 
$(\rep{20},\rep{4},\rep{1}) + (\rep{20},\rep{1},\rep{2})$ & 
$(\rep{1},\crep{4},\rep{2}) + (\rep{1},\rep{6},\rep{1}) +
(\rep{1},\rep{1},\rep{1})$ &  
$\arry{c}{(\rep{1},\crep{4},\rep{2}) + (\rep{1},\rep{6},\rep{1}) \\ 
+ 3(\rep{1},\rep{1},\rep{1})}$ 
\\[1ex]\hline &&&& \\ [-2ex]
5b & $\SO{20}\times\U{3}\times\U{1}^3$ & 
$3(\rep{20},\rep{1}) + (\rep{20},\rep{3})$ & 
$3(\rep{1},\crep{3}) + (\rep{1},\rep{3}) +
3 (\rep{1},\rep{1},\rep{1})$ &  
$2(\rep{1},\crep{3}) +  5 (\rep{1},\rep{1})$ 
\\[1ex]\hline &&&& \\ [-2ex]
6 & $\SO{18}\times\U{4}\times\U{2}\times\U{1}$ & 
$\arry{c}{(\rep{18},\rep{4},\rep{1}) + (\rep{18},\rep{1},\rep{2})\\
2(\rep{1},\rep{4},\rep{1}) +2(\rep{1},\rep{1},\rep{2})}$ & 
$(\rep{1},\crep{4},\rep{2}) + (\rep{1},\rep{6},\rep{1}) +
(\rep{1},\rep{1},\rep{1})$ &  
$\arry{c}{2(\rep{1},\crep{4},\rep{1}) + (\rep{18},\rep{1},\rep{1}) \\ 
+2 (\rep{1},\rep{1},\rep{2})+ (\rep{1},\rep{1},\rep{1})}$ 
\\[1ex]\hline &&&& \\ [-2ex]
7a & $\SO{16}\times\U{3}\times \U{2}^2\times\U{1}$ & 
$ \arry{c}{(\rep{16},\rep{1},\rep{1},\rep{1})  + 
(\rep{16},\rep{1},\rep{1},\rep{2}) \\
+ (\rep{16},\crep{3},\rep{1},\rep{1})+
 2(\rep{1},\rep{3},\rep{2},\rep{1}) \\
+ 2  (\rep{1},\rep{1},\rep{2},\rep{2}) + 
2 (\rep{1},\rep{1},\rep{2},\rep{1}) }$ & 
$\arry{c}{(\rep{1},\rep{3},\rep{1},\rep{1}) +  
(\rep{1},\crep{3},\rep{1},\rep{1}) \\
+(\rep{1},\rep{1},\rep{1},\rep{2}) +  
(\rep{1},\crep{3},\rep{1},\rep{2}) \\
(\rep{1},\rep{1},\rep{1},\rep{1})} $ & 
$  2 (\rep{1},\rep{3},\rep{1},\rep{1}) +  (\rep{1},\rep{1},\rep{1},\rep{1}) $
\\[1ex]\hline &&&& \\ [-2ex]
7b & $\SO{16}\times\U{2}\times \U{5}\times\U{1}$ & 
$ \arry{c}{(\rep{16},\rep{1},\rep{5})  + (\rep{16},\rep{1},\rep{1}) \\
+ 2  (\rep{1},\rep{2},\crep{5}) + 2  (\rep{1},\rep{2},\rep{1}) }$ & 
$  (\rep{1},\rep{1},\rep{10}) +  (\rep{1},\rep{1},\rep{5}) $ & 
$  2 (\rep{1},\rep{1},\crep{5}) +  (\rep{1},\rep{1},\rep{1}) $
\\[1ex]\hline &&&& \\ [-2ex]
8 & $\SO{12} \times \U{4} \times \U{2} \times \U{4}$ & 
$ \arry{c}{(\rep{12},\rep{1},\rep{2},\rep{1}) +
(\rep{12},\rep{4},\rep{1},\rep{1}) \\ + 
(\rep{1},\crep{4},\rep{1},\rep{4}) +
(\rep{1},\crep{4},\rep{1},\crep{4}) \\ + 
(\rep{1},\rep{1},\rep{2},\rep{4}) + 
(\rep{1},\rep{1},\rep{2},\crep{4}) }$ &
$  \arry{c}{ (\rep{1},\rep{6},\rep{1},\rep{1}) +
(\rep{1},\crep{4},\rep{2},\rep{1}) \\ +
(\rep{1},\rep{1},\rep{1},\rep{1}) }$ &  
$  (\rep{1},\rep{1},\rep{1},\rep{6}) +  
(\rep{1},\rep{1},\rep{1},\rep{1}) $
\\[1ex]\hline
\end{tabular}
\end{center}
\captn{\label{tb:spectraC3Z4}
This table gives the chiral part of the spectrum of the resolution
models of the $\Cplx^3/\Intr_4$ orbifold. The models, defined by the
blowup vectors, $V_1$ and $V_2\,$, are numbered according to the
convention defined in table~\ref{tb:modelsC3Z4}.  
}
\end{table}
\end{landscape} 
}

\def\spectraCthreeZfourbis{
\thispagestyle{empty} 
\begin{landscape}
\begin{table}
\begin{center}
\begin{tabular}{| c || c | c | c | c |}
\hline &&&& \\ [-2ex]
Nr. & 4D gauge group & $\frac 18 \times$ ``untwisted''  & 
$\frac 14 \times$ ``2nd twisted'' & ``1st twisted'' 
\\[1ex]\hline &&&& \\ [-2ex]
9 & $\U{5}\times \U{9} \times \U{1}^2$ & 
$ \arry{c}{ (\rep{5},\rep{9}) + (\crep{5},\rep{9}) +
(\rep{5},\rep{1}) \\ +(\crep{5},\rep{1}) + 2 (\rep{1},\crep{9}) + 2
(\rep{1},\rep{1}) }$ & 
$  (\rep{10},\rep{1}) + (\crep{5},\rep{1}) $ & 
$ (\rep{1},\crep{9}) + 2 (\rep{1},\rep{1}) $ 
\\[1ex]\hline &&&& \\ [-2ex]
10 & $\U{3} \times \U{10} \times \U{2} \times \U{1}$ & 
$  \arry{c}{(\rep{3},\rep{10},\rep{1}) +  (\crep{3},\rep{10},\rep{1})
  + \\  
2  (\rep{1},\crep{10},\rep{2}) + 2  (\rep{1},\crep{10},\rep{1}) }$ &
$ \arry{c}{2 (\crep{3},\rep{1},\rep{1}) +  (\crep{3},\rep{1},\rep{2}) 
\\ +  (\rep{1},\rep{1},\rep{2}) +   (\rep{1},\rep{1},\rep{1}) }$ &
$  (\rep{3},\rep{1},\rep{1}) +  (\rep{1},\rep{1},\rep{2}) $ 
\\[1ex]\hline &&&& \\ [-2ex]
11 & $\U{13}\times \U{1}^3$
& $4(\rep{13})+4(\rep{1})$
& $2(\crep{13})+5(\rep{1})$
& $2(\rep{1})$
\\[1ex]\hline &&&& \\ [-2ex]
12a & $\U{13}\times \U{2} \times \U{1}$
& $2(\rep{13},\rep{2})+2(\rep{1},\rep{2})$
& $2(\rep{13},\rep{1})+2(\rep{1},\rep{2})+(\rep{1},\rep{1})$
& $(\crep{13},\rep{1})$
\\[1ex]\hline &&&& \\ [-2ex]
12b & $\U{12}\times \U{2} \times \U{1}^2$
& $2(\rep{12},\rep{2})+4(\rep{1},\rep{2})$
& $2(\rep{12},\rep{1})+2(\rep{1},\rep{2})+3(\rep{1},\rep{1})$
& $(\crep{12},\rep{1})+3(\rep{1},\rep{1})$
\\[1ex]\hline &&&& \\ [-2ex]
13a & $\U{12}\times \U{2} \times \U{1}^2$
& $\arry{c}{(\rep{66},\rep{1})+(\rep{12},\rep{1})
+(\crep{12},\rep{1})\\+(\crep{12},\rep{2})
+2(\rep{1},\rep{2})+2(\rep{1},\rep{1})}$
& $(\rep{12},\rep{1})+(\rep{1},\rep{2})+(\rep{1},\rep{1})$
&  $(\crep{12},\rep{1})+2(\rep{1},\rep{2})+3(\rep{1},\rep{1})$
\\[1ex]\hline &&&& \\ [-2ex]
13b & $\U{13}\times \U{2} \times \U{1}$
& $\arry{c}{(\rep{78},\rep{1})+(\crep{13},\rep{2})\\
+(\crep{13},\rep{1})
+(\rep{1},\rep{2})+(\rep{1},\rep{1})}$
& $(\rep{13},\rep{1})+(\rep{1},\rep{2})$
&  $(\crep{13},\rep{1})+2(\rep{1},\rep{2})+2(\rep{1},\rep{1})$
\\[1ex]\hline &&&& \\ [-2ex]
13c & $\U{11}\times \U{3} \times \U{1}^2$
& $\arry{c}{(\crep{55},\rep{1})+(\rep{11},\rep{3})\\
+2(\crep{11},\rep{1}) +3(\rep{1},\rep{3})+(\rep{1},\rep{1})}$
& $(\crep{11},\rep{1})+(\rep{1},\rep{3})+(\rep{1},\rep{1})$
&  $(\rep{11},\rep{1})+2(\rep{1},\crep{3})$
\\[1ex]\hline &&&& \\ [-2ex]
14a & $\U{5} \times \U{9}\times \U{1}^2$ & 
$ \arry{c}{ (\rep{10},\rep{1}) + 2 (\rep{5},\rep{1}) +
(\crep{5},\crep{9}) \\ + 2 (\rep{1},\crep{9}) +  (\rep{1},\rep{36})
+(\rep{1},\rep{1}) }$ & 
$ (\crep{5},\rep{1}) + (\rep{1},\rep{9}) + (\rep{1},\rep{1}) $ & 
$ (\rep{5},\rep{1}) $
\\[1ex]\hline &&&& \\ [-2ex]
14b & $\U{6} \times \U{8}\times \U{1}^2$ & 
$ \arry{c}{ (\crep{15},\rep{1}) +  (\crep{6},\rep{1}) +
(\rep{6},\rep{1}) \\ +  (\rep{6},\crep{8}) + (\rep{1},\rep{8}) 
+ (\rep{1},\crep{8}) \\ + (\rep{1},\rep{28})+(\rep{1},\rep{1})
}$ 
& 
$ (\rep{6},\rep{1}) + (\rep{1},\rep{8}) + (\rep{1},\rep{1}) $ 
& 
$ (\crep{6},\rep{1})+(\rep{1},\rep{1}) $
\\[1ex]\hline &&&& \\ [-2ex]
15 & $\U{10} \times \U{3}\times \U{2}\times \U{1}$ 
& $\arry{c}{(\rep{45},\rep{1},\rep{1})+(\rep{10},\rep{1},\rep{1})\\
+(\crep{10},\crep{3},\rep{1})+(\crep{10},\rep{1},\rep{2})\\
+2(\rep{1},\crep{3},\rep{1})+(\rep{1},\rep{3},\rep{2})\\
+(\rep{1},\rep{1},\rep{2})+(\rep{1},\rep{1},\rep{1})}$
& $\arry{c}{(\crep{10},\rep{1},\rep{1})\\+
(\rep{1},\rep{3},\rep{1})+(\rep{1},\rep{1},\rep{2})}$
& $\arry{c}{(\rep{1},\crep{3},\rep{1})\\+
(\rep{1},\rep{3},\rep{2})+2(\rep{1},\rep{1},\rep{1})}$
\\[1ex]\hline &&&& \\ [-2ex]
16a & $\U{13}\times \U{1}^3$
& $(\rep{78})+2(\rep{13})+(\crep{13})+3(\rep{1})$
& $(\crep{13})+2(\rep{1})$
& $(\crep{13})+4(\rep{1})$
\\[1ex]\hline &&&& \\ [-2ex]
16b & $\U{14}\times \U{1}^2$
& $(\rep{91})+(\rep{14})+(\crep{14})+(\rep{1})$
& $(\crep{14})+(\rep{1})$
& $(\crep{14})+3(\rep{1})$
\\[1ex]\hline
\end{tabular}
\end{center}
\captn{\label{tb:spectraC3Z4bis}
This table gives the continuation of table~\ref{tb:spectraC3Z4}.
}
\end{table}
\end{landscape} 
\resettablenr
}

\def\intersectionsCthreeZtwoSqsym{
\begin{table} 
\begin{center} 
\begin{tabular}{| c || c | c | c | c | c | c |}
\hline &&&&&& \\ [-2ex]
  & $D_1$ & $D_2$ & $D_3$ & $E_1$ & $E_2$ & $E_3$ 
\\[1ex]\hline\hline &&&&&& \\ [-2ex]
$E_1 E_2$ & $0$ & $0$ & $1$ & $-1$ & $-1$ & $1$ 
\\[1ex]\hline &&&&&& \\ [-2ex]
$E_1 E_3$ & $0$ & $1$ & $0$ & $-1$ & $1$ & $-1$ 
\\[1ex]\hline &&&&&& \\ [-2ex]
$E_2 E_3$ & $1$ & $0$ & $0$ & $1$ & $-1$ & $-1$ 
\\[1ex]\hline\hline &&&&&& \\ [-2ex]
$D_1 E_1$ & $0$ & $-\frac 12$ & $-\frac 12$ & $1$ & $0$ & $0$ 
\\[1ex]\hline &&&&&& \\ [-2ex]
$D_2 E_2$ & $-\frac 12$ & $0$ & $-\frac 12$ & $0$ & $1$ & $0$ 
\\[1ex]\hline &&&&&& \\ [-2ex]
$D_3 E_3$ & $-\frac 12$ & $-\frac 12$ & $0$ & $0$ & $0$ & $1$ 
\\[1ex]\hline
\end{tabular} 
\end{center}
\captn{\label{tb:intersectionsC3Z22sym}
The upper part of the table gives the intersection numbers of the
compact curves with all divisor of the ``symmetric'' resolution of
$\Cplx^3/\Intr_2 \times \Intr_2'\,$. The lower part gives the values
of the integrals of the divisors corresponding to curves, that are not
realized in the symmetric resolution. 
}
\end{table}
}

\def\intersectionsCthreeZtwoSqE{
\begin{table} 
\begin{center} 
\begin{tabular}{| c || c | c | c | c | c | c |}
\hline &&&&&& \\ [-2ex]
  & $D_1$ & $D_2$ & $D_3$ & $E_1$ & $E_2$ & $E_3$ 
\\[1ex]\hline\hline &&&&&& \\ [-2ex]
$E_1 E_2$ & $1$ & $0$ & $1$ & $0$ & $-2$ & $0$ 
\\[1ex]\hline &&&&&& \\ [-2ex]
$E_1 E_3$ & $1$ & $1$ & $0$ & $0$ & $0$ & $-2$ 
\\[1ex]\hline &&&&&& \\ [-2ex]
$D_1 E_1$ & $-1$ & $-\frac 12$ & $-\frac 12$ & $0$ & $1$ & $1$ 
\\[1ex]\hline\hline &&&&&& \\ [-2ex]
$E_2 E_3$ & $0$ & $0$ & $0$ & $0$ & $0$ & $0$ 
\\[1ex]\hline &&&&&& \\ [-2ex]
$D_2 E_2$ & $-\frac 12$ & $0$ & $-\frac 12$ & $0$ & $1$ & $0$ 
\\[1ex]\hline &&&&&& \\ [-2ex]
$D_3 E_3$ & $-\frac 12$ & $-\frac 12$ & $0$ & $0$ & $0$ & $1$ 
\\[1ex]\hline
\end{tabular} 
\end{center}
\captn{\label{tb:intersectionsC3Z22E1}
The upper part of the table gives the intersection numbers of the
compact curves with all divisor of the ``$E_1$'' resolution of
$\Cplx^3/\Intr_2 \times \Intr_2'\,$. The lower part gives the values
of the integrals of the divisors corresponding to curves, that are not
realized in the ``$E_1$'' resolution. 
}
\end{table}
}

\def\modelsCthreeZtwoSqsym{
\begin{table} 
\begin{center} 
\begin{tabular}{| c | c || c | c | c |}
\hline &&&& \\ [-2ex]
orbifold & orbifold & blowup & blowup & blowup 
\\ [-1ex] &&&& \\[-2ex] 
shift $2\,v_1$ & shift $2\,v_2$ & vector $V_1$ & vector $V_2$ & vector
$V_3$
\\[1ex]\hline\hline &&&& \\ [-2ex]
$(1^2,0^{14})$ & $(0,1^2, 0^{13})$ & 
$(1^2,0,2, 0^{12})$ & $(0,1^2,0,2, 0^{11})$ &
 $(1,0,1,0,0,2, 0^{10})$ 
\\[1ex]
&& $(1^2,2,0^{13})$ & $(0,-1,1,2,0^{12})$ &
 $(-1,0,1,0,2, 0^{11})$ 
\\[1ex]\hline &&&& \\ [-2ex]
$(1^2,0^{14})$ & $(0, 1^6, 0^{9})$ & 
$(1^2, 0^{13},2)$ & $(0, 1^6, 0^{9})$ & 
$(1,0, 1^3,-1^2, 0^{9})$  
\\[1ex] &&
$(1^2,2, 0^{13})$ & $(0, -1,1^5, 0^{9})$ & 
$(-1,0, 1^3,-1^2, 0^{9})$  
\\[1ex]\hline &&&& \\ [-2ex]
$(1^6,0^{10})$ & $(0^3,1^6, 0^{7})$ & 
$(1^6, 0^{10})$ & $(0^3,-1,1^5, 0^{7})$&
$(1^2,-1,0^3,1^2,-1, 0^{7})$  
\\[1ex]\hline &&&&\\ [-2ex]
$(1^6,0^{10})$ & $(0^5,1^6, 0^{5})$ & 
$(1^6, 0^{10})$ & $(0^5,1^6, 0^{5})$&
$(0^5,1,0^5,1^5)$ 
\\[1ex]\hline &&&& \\ [-2ex]
$(1^2,0^{14})$ & $\frac{1}{2}(1^{15}, -3)$ & 
$(-1,1,2, 0^{13})$ & $\frac{1}{2} (1^{15}, -3)$ &
$\frac{1}{2}  (-1^2,1^{12}, -3,1)$ 
\\[1ex]\hline &&&& \\ [-2ex]
$(1^6,0^{10})$ & $\frac{1}{2}(-3,1^{15})$ & 
$(1^6, 0^{10})$ & $\frac{1}{2}(-3,1^{15})$&
$\frac{1}{2}(-3,1^5,-1^{10})$  
\\[1ex] &&
$(1^4,-1^2, 0^{10})$ & $\frac{1}{2}(1^{15},-3)$&
$\frac{1}{2}(1^6,-1^8,3,-1)$  
\\[1ex]\hline
\end{tabular} 
\end{center}
\captn{\label{tb:modelsC3Z22sym}
This table compares the $\Cplx^3/\Intr_2\times \Intr_2'$ orbifold
gauge shift vectors, $v_2$ and $v_3\,$, with the blowup vectors,
$V_1\,$, $V_2\,$, and $V_3\,$, that characterize gauge background of
the symmetric resolution of this orbifold. The blowup vectors satisfy
all the flux quantization conditions~\eqref{quantfluxz2z2} and all the
Bianchi identities~\eqref{6DBianchiC3Z22} and~\eqref{4DBianchiC3Z22}.
The identification of the orbifold and blowup shifts is performed upto
lattice vectors.
}
\end{table}
}

\begin{document}

\thispagestyle{empty}

\begin{flushright}
HD-THEP-07-18 \\
SIAS-CMTP-07-4 
\end{flushright}
\vskip 2 cm
\begin{center}
{\Large {\bf Toric Resolutions of Heterotic Orbifolds} 
}
\\[0pt]

\bigskip
\bigskip {\large
{\bf Stefan Groot Nibbelink$^{a,b,}$\footnote{
{{ {\ {\ {\ E-mail: grootnib@thphys.uni-heidelberg.de}}}}}}},
{\bf Tae--Won\ Ha$^{a,}$\footnote{
{{ {\ {\ {\ E-mail: tha@tphys.uni-heidelberg.de}}}}}}},
{\bf Michele Trapletti$^{a,}$\footnote{
{{ {\ {\ {\ E-mail: M.Trapletti@thphys.uni-heidelberg.de}}}}}}}
\bigskip }\\[0pt]
\vspace{0.23cm}
${}^a$ {\it 
Institut f\"ur Theoretische Physik, Universit\"at Heidelberg, \\ 
Philosophenweg 16 und 19,  D-69120 Heidelberg, Germany 
\\} 
\vspace{0.23cm}
${}^b$ {\it 
Shanghai Institute for Advanced Study, 
University of Science and Technology of China,\\ 
99 Xiupu Rd, Pudong, Shanghai 201315, P.R.\ China
 \\} 
\bigskip
\end{center}

\subsection*{\centering Abstract}

We investigate resolutions of heterotic orbifolds using toric
geometry. Our starting point is provided by the recently constructed
heterotic models on explicit blowup of $\Cplx^n/\Intr_n$
singularities. We show that the values of the relevant integrals, computed
there, can be obtained as integrals of divisors (complex codimension
one hypersurfaces) interpreted as $(1,1)$--forms in toric 
geometry. Motivated by this we give a self contained 
introduction to toric geometry for non--experts, focusing on those
issues relevant for the construction of heterotic models on toric
orbifold resolutions. We illustrate the methods by building heterotic
models on the resolutions of $\Cplx^2/\Intr_3\,$, 
$\Cplx^3/\Intr_4$ and $\Cplx^3/\Intr_2\times\Intr_2'\,$. We are able
to obtain a direct identification between them and the known
orbifold models. In the $\Cplx^3/\Intr_2\times\Intr_2'\,$ case we
observe that, in spite of the existence of two inequivalent
resolutions, fully consistent blowup models of heterotic orbifolds can
only be constructed on one of them.

\newpage 
\setcounter{page}{1}

\section{Introduction}
\labl{sc:intro}

One of the main aims of string phenomenology is to build a string model
reproducing, at low energies, the standard model of particle physics,
or a supersymmetric extension of it. 
This issue has been faced from different perspectives, in particular
we remind the reader of models built using  free--fermion
models~\cite{Faraggi:1989ka,Faraggi:1991jr,Cleaver:1998sa}, 
intersecting D--branes in type II string theory~\cite{Berkooz:1996km,
Blumenhagen:2000wh,Aldazabal:2000dg,Honecker:2004kb},
Gepner models~\cite{Dijkstra:2004cc,Dijkstra:2004ym}, 
and compactifications of the heterotic string.
In the latter case, in order to obtain four dimensional
models with at most $N=1$ supersymmetry, i.e. in order to have a chiral
spectrum, one needs to compactify on a Calabi--Yau
space~\cite{Candelas:1985en} (see also \cite{Andreas:1999ty,
Braun:2005ux,Braun:2005bw,Braun:2005nv,Blumenhagen:2005zg,
Blumenhagen:2006ux} for recent progresses in this direction), or on a
singular limit of it: an orbifold. Orbifolds are particularly
convenient, since they allow fully calculable string
compactifications, in terms of combinations of free conformal field
theories~\cite{dixon_85,Dixon:1986jc}. Given this calculability, it is
possible to produce a vast but controllable landscape of models, and
scan among them for realistic ones. Indeed, this approach has been
proven to be successful, and models extremely close to the MSSM have
been
built~\cite{Forste:2004ie,Kobayashi:2004ya,Kobayashi:2004ud,Buchmuller:2004hv,Lebedev:2006kn,Lebedev:2006tr,Kim:2006hw,Kim:2007mt}.

Orbifolds are special points in the full moduli space of the heterotic
string on Calabi--Yau manifolds. In order to have a better control on
the theory away from these special orbifold points, it is crucial to
have a better understanding of model building on the corresponding
smooth compactification spaces. As the theory is completely calculable
at the orbifold point, one may also hope, that one can learn about its 
properties in the moduli space in the vicinity of this singularity. 
The underlying theme of this paper is precisely to study the interplay
between the heterotic string theory at the orbifold points of the moduli
space and on generic Calabi--Yau spaces.

A concrete way to probe the moduli space surrounding 
orbifold points is to consider blowups of
orbifold singularities in an effective supergravity coupled to super
Yang--Mills description. The idea is to first study the resolution of
isolated singularities and after that obtain a description of a
compact Calabi--Yau by gluing various orbifold resolutions
together. The construction of explicit blowups is unfortunately
not easy. The most known example is the Eguchi--Hanson
resolution~\cite{Eguchi:1978xp} of the $\Cplx^2/\Intr_2$ orbifold
singularity. Generalizations to $\Cplx^n/\Intr_n$ were discussed in
the mathematical literature~\cite{Calabi:1979}, see
also~\cite{pol_2,Joyce:2000}. The construction and the application of
explicit blowups of these singularities to heterotic model building has
been investigated in~\cite{Ganor:2002ae,Nibbelink:2007rd}. In
particular, it was shown that all $\Cplx^2/\Intr_2$ and
$\Cplx^3/\Intr_3$ heterotic orbifold models could be recovered by
considering $\U{1}$ bundle gauge backgrounds on the
blowup~\cite{Nibbelink:2007rd}. This construction was used to
explicitly verify, that in blowup multiple anomalous $\U{1}$'s
are possible~\cite{Blumenhagen:2005ga,Blumenhagen:2005pm}, even though
heterotic orbifold models always have at most a single
anomalous $\U{1}\,$. The way out of this apparent paradox is, that a
twisted state, with a non--vanishing VEV, can be reinterpreted as 
a model dependent axion, that can cancel non--universal
anomalies~\cite{GrootNibbelink:2007ew}. This in particular helped to
resolve
confusion~\cite{Kakushadze:1997wx,Kakushadze:1998cd,Lalak:1999bk}
concerning the heterotic/type--I duality on $\Intr_3$ orbifolds.

Explicit blowups of $\Cplx^n/\Intr_n$ singularities were possible,
because both these orbifolds and their blowups have a large isometry
group. However, for four dimensional string model building, these
blowups can only be used to model $\Cplx^2/\Intr_2$ and
$\Cplx^3/\Intr_3$ singularities, while MSSM like model building
seems to require more complicated orbifolds, like $T^6/\Intr_{6-II}$ or
$T^{6}/\Intr_{12-I}$. (See e.g.~\cite{Forste:2004ie,Kobayashi:2004ya,Kobayashi:2004ud,Buchmuller:2004hv,Lebedev:2006kn,Lebedev:2006tr,Kim:2006hw,Kim:2007mt}.)
The singularities of these orbifolds are more complicated and might not
allow for a simple explicit blowup construction. On the other
hand, the topological properties of such resolutions can be
conveniently described by toric geometry, see
e.g.~\cite{Erler:1992ki}. In this paper we explain how using toric 
geometry one can construct heterotic models on resolutions on
arbitrarily complicated orbifold singularities.

For a general mathematical introduction to the subject of toric
geometry we refer the reader to
e.g.\cite{Fulton,Oda,Cox:1993fz,Hori:2003ic,Bouchard:2007ik}.
Applications of toric geometry to orbifold resolutions have also
recently been discussed in~\cite{Lust:2006zh,Reffert:2007im}.  The
presentation of the toric geometry in this paper gives an
exposition of simple toric techniques, which can be used
to understand the topological properties relevant for model building. 
For this program we explain the construction of toric varieties, that
represent the resolution of orbifolds. The divisors, complex
codimension one hypersurfaces, encode the topology of the
resolution. We explain, that one can interpret divisors as
$(1,1)$--forms, and integrate them over the resolution. 
This allows us to use divisors as field strengths, i.e.\ first Chern
classes, of $\U{1}$ complex line bundle gauge backgrounds. These 
backgrounds can then be used to construct consistent heterotic models
on the resolution. To crosscheck this procedure we first reproduce all
results obtained using the explicit blowup of 
$\Cplx^n/\Intr_n\,$. After that we extend the analysis to more 
complicated orbifolds, for which to our knowledge no explicit blowup
has been written down.

To present our results the paper has been structured as follows: In
section~\ref{sc:explicitblowups} we first review the explicit blowup
of the $\Cplx^n/\Intr_n$ orbifold. After that we introduce toric
geometrical techniques to re--obtain the integrals computed on the
explicit blowup as integrals of certain divisors over the
corresponding toric variety. In
section~\ref{sc:multipleexceptionaldivisors} we first give  
a general account of the analysis of orbifold singularities using
toric geometry, and explain how this can be applied to heterotic model
building on orbifold resolutions. We illustrate the various methods by
two examples: The resolution of $\Cplx^2/\Intr_3\,$, the simplest
example of blowup with two exceptional divisors, is
described in subsection~\ref{sc:C2Z3}. The next subsection is devoted
to the resolution of $\Cplx^3/\Intr_4\,$. For both these resolutions
we explain how we can construct consistent models on them, and derive
the conditions, that ensure they have a direct orbifold interpretation
as well. For the $\Cplx^3/\Intr_4$ resolution we construct models that
satisfy possible Bianchi identities, and we confirm, that they give
rise to models free of non--Abelian anomalies in four dimension, which
all can be matched to heterotic orbifolds. Section~\ref{sc:multipleresolutions}
investigates orbifolds that do not possess a single unique
resolution. We propose minimal requirements of defining integrals
avoiding inconsistencies with the linear equivalence relations. The issues,
that arise when the resolution is not unique, are exemplified by
discussing the two inequivalent resolutions of $\Cplx^3/\Intr_2\times
\Intr_2'$ in subsection~\ref{sc:C3Z22}. In the final
subsection~\ref{sc:hetC3Z22} we compute heterotic models on one of the
resolutions, and argue that no fully consistent model can be built on
the other. In section~\ref{sc:concl} we summarize our conclusions.

\section{Toric description of explicit blowups of orbifold singularities}
\labl{sc:explicitblowups}

\subsection{Blowup of $\Cplx^n/\Intr_n$ orbifold}
\labl{sc:blowupCnZn}

In~\cite{Nibbelink:2007rd} we have given a detailed
description of how to explicitly obtain a blowup of the
$\Cplx^n/\Intr_n$ orbifold with possible $\U{1}$ bundles. Here we will
only recall those results which will be relevant for our subsequent 
discussion, for details the reader may
consult~\cite{Nibbelink:2007rd,Ganor:2002ae}. The $\Cplx^n/\Intr_n$
orbifold is defined by the $\Intr_n$ action 
\equ{
\gTh (\tZ) ~=~ \gth \, \tZ~, 
\qquad 
\gth ~=~ e^{2\pi i\, \gf}~, 
\qquad 
\gf ~=~ \frac 1n\, \text{diag} 
\Big( 1, \ldots, 1  \Big)~,
\labl{geomshift}
}
on the orbifold coordinates $\tZ\,$. This defines a space with a
singularity, having deficit angle of  $2\gp (1-1/n)\,$. 
The geometry of the non--singular blowup is described
by the \Kh\ potential $\cK$ given by
\equ{
\cK(X) ~=~ 
\int\limits_1^X \frac {\d X'}{X'} \, M(X')~, 
\qquad 
M(X) ~=~ \frac 1{n} \big( r + X \big)^{\frac 1n}~. 
\labl{FunK}
}
where $X = (1 + \bz z)^n |x|^2$ is an $\SU{\mbox{$n$}}$ invariant, and
the $z$ and $x$ are the coordinates of the space. In detail, 
the $z$ form a set of inhomogeneous complex coordinates of $\CP^{n-1}\,$,
and $x$ the  coordinate parameterizing the complex line over
$\CP^{n-1}\,$. Finally, $r$ is the resolution parameter, defined such
that in the limit $r \ra 0$ one retrieves the orbifold geometry.

From the \Kh\ potential all geometrical quantities can be derived in
the standard way, in particular, the curvature 2--form reads 
\equ{
\cR ~=~ \frac{r}{r + X} \, 
\pmtrx{ \dsp 
 e \, \bee ~-~ \bee \, e  
~+~ \frac 1{n}\, \frac{ \bge \, \ge}{r + X}
& \dsp 
\frac{\bge \, e} {\sqrt{r + X}} 
\\[3ex] \dsp 
\frac{\bee\, \ge}{\sqrt{r + X}} 
& \dsp 
n\, \bee \, e ~-~ \frac {n\!-\!1}{n} \,  \frac{ \bge \, \ge}{r + X}
}~.
\labl{Curv2expl}
}
Here $e$ and $\ge$ are the holomorphic vielbein 1--forms of
$\CP^{n-1}$ and its complex line bundle. It can be shown that $\cR$ is
traceless, which is consistent with the Calabi--Yau property of having
vanishing first Chern class. In addition, this geometry admits a
$\U{1}$ gauge background, that satisfies the Hermitian Yang--Mills
equations, with field strength 2--form 
\equ{
i\cF ~=~ \Big(\frac r{r + X} \Big)^{1-\frac 1n}
\Big( 
\bee e ~-~ \frac {n-1}{n^2} \, \frac 1{r+X}\, \bge \ge
\Big)~. 
\labl{FU1basis}
}

Because both the geometry and its $\U{1}$ gauge background are given
explicitly, integrals of them can be computed straightforwardly. In
particular, we obtain
\equ{ 
\int_{\CP^2} \frac{\tr\, \cR^2}{(2\gp i)^2} 
~=~ -n \, 
 \int_{\CP^1\ltimes \Cplx}  \frac {\tr\, \cR^2}{(2\gp i)^2}
~=~ 
n(n+1)~, 
\labl{trR2int}
} 
and 
\equ{
\int_{\CP^p} 
\Big( \frac { i \cF}{2\gp i} \Big)^p
~=~ 
- n
\int_{\CP^{p-1} \ltimes \Cplx} 
\Big( \frac { i \cF}{2\gp i} \Big)^p
~=~ 
1~. 
\labl{trFpint}
}
The integrals over $\CP^{p}$ are taken at $X=0$ integrating over $p$
of the $n-1$ inhomogeneous coordinates of $\CP^{n-1}$, with the others
set to a fixed value, say, $0$. The integral over 
$\CP^{p-1} \ltimes \Cplx$ corresponds to the integral over all values
of  $x \in \Cplx$ and over $p-1$ inhomogeneous coordinates.

These and other integrals were relevant to determine the heterotic
blowup models that satisfy the integrated version of the Bianchi
identity
\equ{
\d H ~=~ \tr\, \cR^2 \,-\, \tr (i\cF_V)^2~, 
}
where $i\cF_V = i \cF\, V^I H_I$ defines the embedding of the $\U{1}$
gauge background in the $\SO{32}$ or $\E{8}\times\E{8}$ gauge group. 
Integrating the Bianchi identity over the full blowup of
$\Cplx^2/\Intr_2$ and requiring that it vanishes, leads to the
consistency condition $V^2 = 6\,$. In the three dimensional case 
the integral in the Bianchi identity over either $\CP^2$ or 
$\CP^1\ltimes \Cplx$ lead to the same consistency condition 
$V^2 = 12\,$ for the blowup of $\Cplx^3/\Intr_3\,$. Both conditions in
two and three complex internal dimensions are compatible with the
corresponding modular invariance conditions, 
$(2v)^3 = 2~\text{mod}~4$ and $(3v)^3 = 0~\text{mod}~6\,$, of the
heterotic string, respectively.

Moreover, in~\cite{Nibbelink:2007rd} we confirmed that the integral
or half--integral solutions of this equation, gives rise to all
blowups of all of the known modular invariant $T^4/\Intr_2$ and
$T^6/\Intr_3$ heterotic orbifold models (except the $\Intr_3$ models
with unbroken $\SO{32}$ and $\E{8}\times\E{8}$ gauge groups). 
We identified the gauge background $\cF_V$ with the orbifold
action on the gauge degrees of freedom 
$\fA(\gth \tZ) = U \fA(\tZ) U\inv\,$, with $U = \exp(2\pi i \, v^I H_I)$
characterized by $v^I\,$.  For this we computed the integral over the
contour $\gg$ of the phase of $x$ at $x \ra \infty$ at fixed
values of the $\CP^{n-1}$ coordinates $z$\,: 
\equ{
v^I H_I ~\equiv~ \int_\gg \cA_V ~=~ - \frac 1n \, V^I H_I~. 
\labl{contourInfty}
}
The equivalence sign ``$\equiv$'' indicates, that the identification of
the orbifold gauge shift vector $v\,$, and the blowup parameter $V\,$,
that characterizes the $\U{1}$ bundle embedding in the gauge group, is
upto lattice vectors in the $\text{Spin(32)}$ lattice.

In addition we could use these integrals to compute the complete
chiral spectrum of the blowups using index theorems. We found that the
spectra were identical the orbifold spectra in the blow down limit upto
singlets and vector--like states. The fact that we were able to obtain
the blowups of all heterotic orbifold models and the chiral part of
the spectra, gives us confidence that, even though we are (partly)
integrating over non--compact cycles, the integrals can
nevertheless be trusted and used in a naive way in index
theorems. In particular, we do not have to use extensions of index
theorems for spaces with boundaries, when computing on the blowup of 
non--compact $\Cplx^n/\Intr_n$ orbifolds and comparing this with the
properties of compact $T^{2n}/\Intr_n$ orbifolds. The reason that this
procedure works is, that we in the end compare with the spectrum of a
compact orbifold. This requires, that we glue various resolutions
together. In this process the boundary contributions cancel.

\subsection{Resolution of $\Cplx^n/\Intr_n$ using toric geometry}
\labl{sc:toricCnZn}

The purpose of this subsection is to understand the topology of
the resolution of $\Cplx^n/\Intr_n$ using toric geometry. In
particular, we show how the integrals~\eqref{trR2int} 
and~\eqref{trFpint} can be obtained using this  machinery. Our
description explains the basic methods to obtain the results relevant
for (heterotic) string model building.

As explained below~\eqref{geomshift} the orbifold $\Cplx^n/\Intr_n$
has a deficit angle. To obtain a non--singular resolution
$\Res(\Cplx^n/\Intr_n)$, we define a set of local coordinates 
\equ{
Z_1 ~=~ z_1\, x^{\frac 1n}~,
\quad \ldots \quad  
Z_n ~=~ z_n\, x^{\frac 1n}~,
\labl{localCnZn}
}
from the homogeneous coordinates $z_1,\ldots, z_n, x \in \Cplx\,$. The
orbifold action~\eqref{geomshift} is then extended by the
transformation $x \ra e^{-2\gp i} x\,$. As it stands we describe the
$n$ local coordinates using $n+1$ homogeneous coordinates, we
therefore need to define a $\Cplx^* = \Cplx - {0}$ ``toric'' action on the
homogeneous coordinates, that leave the local coordinates inert. This
requirement fixes the $\Cplx^*$ action uniquely to 
\equ{
\Cplx^*\,: ~ 
(z_1, \ldots, z_n, x) ~\sim~ 
(\gl\inv \,z_1, \ldots, \gl\inv \,z_n, \gl^n \,x)~,
\labl{scalingCnZn} 
}
$\gl \in \Cplx^*\,$. The resolution of $\Cplx^n/\Intr_n$ is defined by
the toric variety 
\equ{
\Res(\Cplx^n/\Intr_n) ~=~ 
\Big( \Cplx^{n+1} \,-\, F \Big)/ \Cplx^*~,
\labl{ResCnZn}
}
where the exclusion set $F$ has been subtracted to ensure, that the
resolution is not singular. In particular, the $\Cplx^*$ action should
act non--trivially, hence at least the origin, 
$\{z_1= \ldots = z_n = x = 0\}\,$, has to be excluded. Indeed, the
number of coordinates set to zero in a toric variety, $p\,$, determines a
subspace of complex dimension $n-p\,$. In particular, one 
expects, that the origin has ``$-1$'' dimensions, and hence
totally irrelevant. But since the $\Cplx^*$ leaves it inert, it is
zero dimensional, i.e., a collection of points, which do matter in
general.

The resolution $\Res(\Cplx^n/\Intr_n)$ is topologically non--trivial,
i.e.\ one needs more than one coordinate patch to describe it
entirely. A set of coordinate patches $U_i$ is obtained
straightforwardly, by taking one of the homogeneous coordinates not to
be vanishing 
\equ{
U_0 ~=~ \big\{\, x ~\neq~0\, \big\}~, 
\quad 
U_i ~=~ \big\{\, z_i ~\neq~0\, \big\}~, 
\label{patchesCnZn}
}
for $i = 1,\ldots n\,$, defined of course in $\Cplx^{n+1}-F$ only. In
each of the coordinate patches we can use the
rescaling~\eqref{scalingCnZn} to set its defining non--vanishing 
coordinates to unity. For $U_i$ this can be done uniquely by setting
$\gl = z_i\,$. But for $U_0$ we find a $\Intr_n$ ambiguity because 
$\gl = e^{2\gp i p/n} x^{-1/n}\,$. Hence on the remaining coordinates
$z_1,\ldots,z_n$ the $\Cplx^*$ reduces to a $\Intr_n$ action. This is
in fact the original orbifold action, and we have a singularity unless
we exclude 
\equ{
F ~=~ \{\,  z_1 = \ldots = z_n = 0 \,\}~.
\labl{exclusionCnZn} 
}
To define proper patches, we need to subdivide the punctured $U_0\,$,
but we will not dwell on this here.

The explicit blowup of the $\Cplx^n/\Intr_n$ orbifold, described in
the previous subsection, used the coordinate patch $U_n$, with 
$z_n=1\,$. In this patch the $\SU{\mbox{$n$}}$ invariant variable $X$ is
obtained from the inhomogeneous coordinates~\eqref{localCnZn}: 
\equ{
X^{1/n} ~=~ \bZ Z ~=~ (1 + \bz z) \big| x^{\frac 1n} \big|^2~. 
}
Only here $z = (z_1, \ldots z_{n-1})$ denote a set of inhomogeneous
coordinates on $\CP^{n-1}\,$. 
The reason, that even though the coordinate patch $U_n$ is not
sufficient to describe the whole resolution, still the integrals give
the correct numbers, is that the parts of $\Res(\Cplx^n/\Intr_n)$ not
in $U_n$ correspond to lower dimensional subspaces, irrelevant for
these integrals.

We define a set of $n+1$ hypersurfaces of complex dimension $n-1\,$, 
which are called divisors. (For a general introduction to algebraic
geometry including divisors see
e.g.~\cite{griffiths78a,huybrechts04a}.) There are two types of divisors, $D_i\,$, 
$i = 1,\ldots n$, and $E$, defined by 
\equ{
D_i ~=~  \big\{\, z_i ~=~0\, \big\}~, 
\qquad 
E ~=~  \big\{\, x ~=~0\, \big\}~.
\labl{divCnZn}
}
The final one, $E\,$, is called an exceptional divisor, because it
defines a subspace of the resolution not present in the
orbifold. Taking into account the remaining
rescaling~\eqref{scalingCnZn}, we see that  $E = \CP^{n-1}$ defined in
terms of homogeneous coordinates. This means that the singularity of
the orbifold $\Cplx^n/\Intr_n$ has been ``blown up'' to a
$\CP^{n-1}\,$. In a similar fashion, it follows that 
$D_i = \CP^{n-2} \ltimes \Cplx$ is defined as a complex line bundle
over $\CP^{n-2}\,$. The resolution $\Res(\Cplx^n/\Intr_n)$ itself can
be thought of as a complex line bundle over $\CP^{n-1}\,$. The
exceptional divisor $E$ is obviously compact, while the other divisors
are not compact.

To each of the divisors we can associate a complex line bundle. Any 
complex line bundle is defined by its holomorphic scalar transition
functions. To determine these transition functions for the various
divisors we write the defining equation of the divisor in
patch $U_i\,$. This gives for the ordinary divisor $D_i\,$: 
\equ{
U_{j\neq i}\,:~~ \frac {z_i}{z_j} ~=~ 0~, 
\quad 
U_{i}\,:~~ 1 ~=~ 0~,
\quad 
U_0\,: ~~ x^{\frac 1n}\,z_i ~=~ 0~, 
}
and for the exceptional divisor $E\,$: 
\equ{
U_{j}\,:~~ z_j^n\,x ~=~ 0~, 
\quad 
U_{0}\,:~~ 1 ~=~0 ~.
}
In the coordinate patches, where we encounter the inconsistent equation
``$1 = 0$'', the corresponding divisor simply does not live. From this
we read off the transition functions for the associated line bundle of
divisors $D_i$ and $E\,$: 
\equ{
g_{kj}(D_i) ~=~ \frac{z_k}{z_j}~, 
\quad 
g_{j0}(D_i) ~=~ x^{\frac 1n}\, z_j~,
\qand 
g_{kj}(E) ~=~ \frac {z_j^n}{z_k^n}~, 
\quad 
g_{j0}(E) ~=~ \frac 1{z_j^n \, x}~. 
}
The subscripts indicate between which two coordinate patches the
transition functions interpolate. It follows, that the transition
functions of the line bundles, associated to the divisors, 
$D_i$ and $E\,$, are all related to each other:  
\equ{
g(D_1)^{-n} ~=~ \ldots ~=~ g(D_n)^{-n} ~=~ g(E)~.
\labl{relTransDivCnZn}
}
Since the equality holds for all transition functions, we have dropped
the subscripts that indicate the coordinate patches.

To understand the consequences of the fact, that all transition
functions of the divisors are related, we make the following brief
excursion to properties of vector bundles. A vector bundle $\cV$ can
be topologically characterized by its total Chern class 
\equ{
c(\cV) ~=~ \det \Big(1 \,+\, \frac{F(\cV)}{2\pi i} \Big)~, 
\labl{totalChern}
}
where $F(\cV)$ is the curvature of the bundle. The total Chern class
can be expanded in terms of its first, second, etc., Chern classes
$c_1(\cV)$, $c_2(\cV)$, etc.  
A complex line bundle is completely determined by its first Chern
class $c_1(\cV) = F(\cV)/2\pi i\,$, which can be taken to be
harmonic (1,1)--form. Because it is closed, locally its
curvature can be written as $F(\cV) = \d A_i(\cV)$ in terms of a
connection $A_i(\cV)$ in coordinate patch $U_i\,$. Between two
coordinate patches $U_i$ and $U_j$ the connections
\equ{
A_j(\cV) ~=~ A_i(\cV) \,+\, g_{ji}(\cV)\inv \,\d\, g_{ji}(\cV)
}
are related via the transition functions $g_{ji}(\cV)\,$.

With this in mind, we can describe the Chern classes of the line
bundles associated to the divisors of the resolution
$\Res(\Cplx^n/\Intr_n)\,$. To each of the divisors
$D_i$ and $E$ of the resolution we can associate a line bundle with
first Chern class, $c_1(D_i)$ and $c_1(E)$, respectively. It is a
convenient  toric geometrical convention, to let the context determine
whether the symbol for the divisor refers to the defining
hypersurface, or the first Chern class of its associated line
bundle. Therefore, one may write $D_i = c_1(D_i)$ and 
$E = c_1(E)\,$. The relations between the transition
functions~\eqref{relTransDivCnZn} imply that the divisors satisfy
the following linear equivalence relations  
\equ{
D_i ~\sim D_j~, 
\qquad 
n\, D_i \,+\, E ~\sim~ 0~, 
\labl{lineqvCnZn}
}
where the linear equivalences, $\sim\,$,  can be replaced by
equalities, provided that the symbols for the divisors refer to the
first Chern classes of the line bundles, when we ignore addition of
exact forms. Upon using Poincar\'e's
duality the divisors refer to hypersurfaces, the linear equivalences
mean, that these surfaces can be deformed to differ by boundary
surfaces. The derivation of the linear equivalence relations
by first determining the relation between the transition
functions~\eqref{relTransDivCnZn} is proper but somewhat lengthy. 
It can be bypassed by requiring that  the local
coordinates~\eqref{localCnZn} are invariant under the transformations 
$z_i \ra e^{D_i}\, z_i$ and $x \ra e^E\, x\,$. The reason that this
works is, that one can perform transformations on the homogeneous
coordinates, that leave the local coordinates~\eqref{localCnZn}
invariant.

The $(1,1)$--forms, $D_i$ and $E\,$, can be integrated over holomorphic 
1--cycles, i.e.\ complex curves. Similarly (2,2)--forms, like
$D_iD_j\,$, $D_i E$ and $E^2\,$, can be integrated over holomorphic
2--cycles, and so on. It is therefore useful to have a classification
of the holomorphic $p$--cycles within the resolution
$\Res(\Cplx^n/\Intr_n)\,$, using the divisors $D_i$ and $E$ interpreted
as hypersurfaces. From their definition it follows immediately that
$D_i$ and $E$ define holomorphic $(n-1)$--cycles. We can define the
integral of any $(n-1,n-1)$--form, say, $D_2^{n-2} E$ over, for example,
$D_1\,$, and denote it by $\int_{D_1} D_2^{n-2} E\,$. 
Moreover, the intersection of two divisors, like
\equ{
D_i \cdot D_{j \neq i} ~=~ 
\big\{\, z_i = z_j = 0 \,\big\}~, 
\qand 
D_i \cdot E ~=~ \big\{\, z_i = x = 0 \, \big\}~,
}
define $(n-2)$--dimensional holomorphic hypersurfaces. The integral
over such intersection of $(n-2,n-2)$--forms can similarly be defined.  
This can of course be extended to the intersection of an arbitrary
number of different divisors. Because $E$ is compact, intersections, that
involve $E\,$, will also be compact; contrary to intersections of only
non--compact divisors $D_i$ can be non--compact. This gives us a way to
identify the integration ranges used in~\eqref{trR2int}
and~\eqref{trFpint}: 
\equ{
\CP^p ~=~ E \, D_1 \ldots D_{n-1-p}~, 
\qquad 
\CP^{p-1}\ltimes \Cplx ~=~ D_1 \ldots D_{n-p}~, 
\labl{cyclesCnZn}
}
with intersections of divisors.

The intersections of $n$ different divisors are of special interest, 
because they define zero dimensional surfaces, i.e.\ sets of
points. The number of such points is called the intersection number of
these divisors. The intersection number of $n-1$ different $D_i$'s and
a single $E$ can be computed directly: 
For example consider $D_2\cdot \ldots \cdot D_n \cdot E$. Setting 
$z_2 = \ldots = z_n = x = 0$ in~\eqref{scalingCnZn}, realizing that 
$z_1 \neq 0\,$, we can choose $\gl = z_1$ uniquely. This means that
all the intersection numbers
\equ{
E \cdot \prod_{j \neq i} D_j  ~=~ 
\int  E D_2 \ldots D_n  ~=~ 1~.
\labl{basicIntCnZn} 
}
The middle equation shows that we can also view these intersection
numbers as integrals over the whole toric variety of the $n$
divisor interpreted as $(1,1)$--forms.

This naturally leads to the following generalization the ``product''
or ``intersection'' of any $n$ divisors can be defined as the integral
over the corresponding $(1,1)$--forms. The linear equivalences to
relate the integral to an integral of all different divisors one of
which being $E\,$. In particular, we find self--intersection number
\equ{
E^n ~=~ (-n)^{n-1} \int D_2 \ldots D_n E ~=~ (-n)^{n-1}~. 
}
In the same way all other (self--)intersections involving at least one
$E$ may be computed. As can be seen from these simple computations 
the symbol $\cdot$ to  indicate intersection of divisors is also
essentially obsolete, and in the following we let the context decide 
whether, say $E D_1$, refers to a $(2,2)$--form  or a complex
$(n-2)$--cycle. Employing the linear equivalence relations we can
even compute integrals over $n$ non--compact divisors, for example 
\equ{
D_1 \ldots D_n ~=~ 
- \frac 1n\, \int E D_2 \ldots D_n ~=~ - \frac 1n~. 
}
This brings us to a few important issues: First of all, one cannot
interpret this result naively as saying that the non--compact divisors
$D_1$ to $D_n$ intersect $-\frac 1n$ times. In fact, the exclusion set
$F$, defined in~\eqref{exclusionCnZn}, implies that this intersection
does not exist in the resolution $\Res(\Cplx^n/\Intr_n)\,$. Hence, one
should {\em only} interpret $D_1 \ldots D_n$ as the integral of the 
corresponding $(1,1)$--forms over the whole resolution.

But even when one interprets $D_1\ldots D_n$ as an integral only, one
may still wonder what fixes its values, because being non--compact it
seems not to be topological. To pursue this question, we explain how
to recover the results for integrals~\eqref{trFpint} using toric
geometry. To obtain the latter integrals we need to identify the gauge
background $i\cF$ with a divisor interpreted as a first Chern class
$(1,1)$--form. The linear equivalences~\eqref{lineqvCnZn} imply that
there is in fact only one independent $(1,1)$--form, hence it is
determined upto an overall normalization. To fix the overall
normalization we look for the $(1,1)$--form which integral is unity on
compact curves, like $E D_2 \ldots D_n\,$, which according
to~\eqref{cyclesCnZn} corresponds to $\CP^n\,$.  In this way we obtain
the identification 
\equ{
\frac{\cF}{2\pi} ~=~ D_i ~=~ - \frac 1n\, E~, 
\qquad 
\int_{E D_2\ldots D_n} \frac{\cF}{2\pi} ~=~ 1~. 
}
Using the identification of the cycles~\eqref{cyclesCnZn} and the
linear equivalences~\eqref{lineqvCnZn} we find the toric formulation
\equ{
\int_{E D_1 \ldots D_{n-1-p}} 
\Big( \frac { i \cF}{2\gp i} \Big)^p
~=~ 
- n
\int_{D_1 \ldots D_{n-p}} 
\Big( \frac { i \cF}{2\gp i} \Big)^p
~=~ 1~, 
}
in agreement with the integrals~\eqref{trFpint}. This shows, that it
is the boundary conditions on $E D_3 \ldots D_n$ or at the boundary of
$D_2 \ldots D_n$ at infinity, which fixes the values of these integrals.
By patching various resolutions together, one can turn
the non--compact divisors and curves into compact ones, and then the
standard intersection theory works, see~\cite{Lust:2006zh}.

Similarly, to obtain a representation of the integrals~\eqref{trR2int}
involving the curvature $\cR\,$, we can employ the splitting
principle~\cite{bott82a},  
which says that the total Chern class $c(\cR)$ of the tangent bundle
is given as the product of $1+D$ over all compact and non--compact
divisors $D\,$. For the resolution of $\Cplx^n/\Intr_n$ this amounts to~\cite{Fulton}
\equ{
c(\cR) ~=~  (1+E) \prod_{i=1}^n (1 + D_i)~. 
\labl{splittingCnZn}
}
The first, second, etc., Chern classes of the tangent bundle can be
determined by expanding this to the appropriate order. As the
resolution represents a (non--compact) Calabi--Yau space, the first
Chern class should vanish. This can be confirmed easily: 
\equ{
c_1(\cR) ~=~ E \,+\, \sum_{i=1}^n D_i ~=~ 0~,  
}
by virtue of the linear equivalence relations~\eqref{lineqvCnZn}. 
By expanding the general formula for the total Chern
class~\eqref{totalChern} to second order gives 
\equ{
- \frac 12\, \tr \Big(\frac{\cR}{2\gp i} \Big)^2 
~=~ c_2(\cR) ~=~ 
E \sum_{i} D_i \,+\, \sum_{i < j} D_i D_j ~=~ 
\frac {n+1}2\, E D_1~, 
\labl{trR2intT}
}
using that the first Chern class vanishes. From this it is
straightforward to confirm the integrals~\eqref{trR2int} of 
$\tr\,\cR^2$ as well.

Next, we want to relate the toric geometry to  heterotic orbifolds. 
In particular we explain how, from it the blowup models characterized
by the vector $V$ of only integers or half--integers, the
corresponding orbifold models defined by the gauge shift $v$ can be 
recovered. The relation between $V$ and $v$ was made
in~\eqref{contourInfty} by computing the contour integral over the 
gauge connection $\cA_V$ far away from the singularity. Using Stoke's
theorem this can be translated to an integral of $\cF_V$ over a  
curve like $D_2 \ldots D_n\,$: 
\equ{
v^I H_I ~\equiv~ \int_{D_2\ldots D_n} \cF_V ~=~ -\frac 1n\, V^I H_I~.
\labl{IdOrbiShiftCnZn}
}
Hence the fractional nature of the orbifold gauge shift vector $v$ is
obtained by integrating over a non--compact curve. 
The integrated version Bianchi Identity is easily computed. For
$\Res(\Cplx^2/\Intr_2)$ we find 
\equ{
V^2 ~=~ -2 \int \tr(\cF_V)^2 ~=~ -2 \int \tr\, \cR^2 ~=~ 6~, 
}
when integrated over the whole resolution. For $\Res(\Cplx^3/\Intr_3)$
we obtain  
\equ{
V^2 ~=~ \int_E \tr (\cF_V)^2 ~=~ -3\, \int_{D_i} \tr(\cF_V)^2
~=~ -3 \int_{D_i} \tr\, \cR^2 ~=~ \int_E \tr\, \cR^2 ~=~ 12~,
}
using~\eqref{trR2intT}. Hence, we have retrieved the conditions
mentioned in the previous subsection. Moreover the toric approach
shows, that the integrals over the compact and non--compact 2--cycles
$E$ and $D_i$ lead to the same condition, is a simple consequence of
the fact, that these divisors are linearly 
equivalent~\eqref{lineqvCnZn}.

\begin{figure}
\begin{center} 
\begin{tabular}{ c c c }
\raisebox{0ex}{\scalebox{0.7}{\mbox{\input{C2Z2.pstex_t}}}}
& \qquad \qquad & 
\raisebox{0ex}{\scalebox{0.5}{\mbox{\input{C3Z3.pstex_t}}}}
\end{tabular} 
\end{center}
\captn{\label{fg:toricCnZn}
The left graph displays the toric diagram of
$\Res(\Cplx^2/\Intr_2)\,$. The right picture displays a projected view
of the toric diagram of $\Res(\Cplx^3/\Intr_3)\,$. Because the latter
is a projection, there are no arrows from the origin pointing to
the divisors as in the former toric diagram. 
}
\end{figure}

There is a convenient way to represent the properties of toric
varieties including the properties of the divisors: the toric
diagram. To build the toric diagram of  
$\Res(\Cplx^n/\Intr_n)$ first give $n$ vectors $v_1, \ldots, v_n$ that
represent the $n$ ordinary divisors $D_1,\ldots, D_n\,$. For example
we can take the basis $v_1 = (1, 0,\ldots, 0)\,$, to 
$v_n = (0, \ldots, 0, 1)\,$. The exceptional divisor $E$ is
represented by the vector  
\equ{
w ~=~ \sum_i \gf_i\, v_i~,
}
which in this basis takes the form $w = (1, \ldots, 1)/n\,$. This
basis $v_1,\ldots, v_n$ and $w$ precisely dictate how to construct the
local coordinates~\eqref{localCnZn}. The toric
diagram of $\Res(\Cplx^2/\Intr_2)$ is given in the left picture of
figure~\ref{fg:toricCnZn}. The toric diagram of
$\Res(\Cplx^3/\Intr_3)$ is three dimensional; to obtain a simple
representation of it we can take a two dimensional projection of the
three dimensional toric diagram. We choose the basis $v_1=(0,0,1)\,$,
$v_2 = (1,0,1)$ and $v_3 = (0,1,1)\,$, so that the exceptional divisor
$E$ is then represented by $w = (\frac 13, \frac 13, 1)\,$. Because
the last entry in both $v_i$ and $w$ are identical, we only need to
use the first two entries, which defines a projection. The
resulting projected toric diagram is given in the right picture in
figure~\ref{fg:toricCnZn}. The exceptional divisor $E$ lies in the
interior of the toric diagram. A theorem in toric geometry guarantees
that such a divisor is compact. We see this theorem confirmed in this 
example. Toric geometry also tells us, that the basic
cones, the smallest possible cones inside a (projected) toric diagram,
correspond to the intersection of divisors with unit intersection
number. This is consistent with~\eqref{basicIntCnZn}, for example, 
$D_1 E =1$ and $D_1 D_2 E = 1\,$, in the resolution,
$\Res(\Cplx^2/\Intr_2)$ and  $\Res(\Cplx^3/\Intr_3)\,$,
respectively. Together with the linear equivalences~\eqref{lineqvCnZn}
we can determine the intersections of a compact curve with the
divisors. We construct the table:  
\begin{center} 
\begin{tabular}{| c | c  c c c |}
\hline &&&& \\ [-2ex]
divisor & $D_1$ & $\ldots$ & $D_n$ & E 
\\[1ex]\hline &&&& \\ [-2ex]
$E D_2\ldots D_n$ & $1$ & $\ldots$ & $1$ & $-n$ 
\\[1ex]\hline
\end{tabular} 
\end{center}
Notice that the values in this table precisely correspond to the minus
the powers of the rescaling parameter $\gl$ in~\eqref{scalingCnZn},
hence we read off the $\Cplx^*$ scaling charges from the toric
diagram, by computing the intersection numbers of a compact curve with
the divisors.

To summarize we have shown that all the results for the integrals
obtained using the explicit blowup of the $\Cplx^n/\Intr_n$ orbifold
singularity can be obtained using toric geometrical techniques, without
ever having to compute any integral explicitly. This procedure shows, 
that the integrals all have a topological origin, which is compatible
with the fact that these integral are used in the integrated Bianchi
identities to select consistent blowup models. All this information
can be obtained uniquely from the toric diagram, which was directly
determined form the orbifold action.

\section{Orbifold resolutions with multiple exceptional divisors}
\labl{sc:multipleexceptionaldivisors}

\subsection{Generalities of orbifold resolutions}
\labl{sc:generalitiesOrbi}

In the previous section we have seen how we can obtain all topological
relevant information of the resolution of $\Cplx^n/\Intr_n$ orbifolds
using toric geometrical techniques. (For related discussions see
e.g.~\cite{Hori:2003ic,Lust:2006zh,Reffert:2007im}.) In this section
we would like to show, that this machinery can be used to treat
resolutions of more complicated orbifolds as well. This requires us to
be able to analyze resolutions with more than one exceptional divisor.

We begin to formalize the toric geometrical method to construct
the resolution of an orbifold singularity by defining the toric
diagram. Consider non--compact orbifolds $\Cplx^n/G\,$, where $G$ is a
finite group, Abelian for simplicity, and $n=2,\,3$.  
The action of an element $\gth
\in G$ on the orbifold coordinates $\tZ_1, \ldots \tZ_n$ can be
written as   
\equ{
\gth\, :~ \big( \tZ_1, \ldots, \tZ_n\big) ~\ra~ 
\big( e^{2\gp i \gf_1(\gth)} \tZ_1, \ldots, 
e^{2\gp i \gf_n(\gth)} \tZ_n \big)~, 
}
such that all $0 \leq \gf_i(\gth) < 1\,$. The elements $\gth$ and
$\gth\inv$ lead to the same orbifold action upto complex conjugation. 
They have to be identified, when $\gth$ acts non--trivially on
three complex dimensions,  but not when it on only acts two complex
coordinates. (A $\Intr_2$ group element $\gth\,$, for which all
$\gf_i(\gth) = 0, 1/2\,$, is self conjugate.) 
We define the corresponding representative $[\gth]$ 
to be the  element that satisfies $\sum_i \gf_i(\gth) = 1\,$. To each 
representative $[\gth]$ we can associate an exceptional divisor
$E_\gth\,$. The total number of exceptional divisors is denoted as
$N\,$. For even and odd ordered orbifolds we encounter 
$N(\Intr_{2k}) = N(\Intr_{2k+1}) = k$ exceptional divisors. 
If we let $v_1, \ldots v_n$ define a basis for the toric
diagram of the orbifold, then the vector 
\equ{
w_\gth ~=~ \sum_i \gf_i(\gth) \, v_i~,
\labl{defw}
}
identifies the exceptional divisor $E_\gth$ in the toric diagram of
the resolution for each representative $[\gth]\,$. This definition of
exceptional divisors of the  
resolution is in one--to--one correspondence to the twisted sectors in
orbifold string theory: Also there each representative $[\gth]$
corresponds to a distinct, e.g.\ first, second, etc., twisted
sectors. In particular,  as is well--known  the $\Cplx^n/\Intr_n$
orbifolds, with $n=2,3\,$, have only a single twisted sector, in
agreements with the previous section where we only had a single
exceptional divisor. The set of vectors $v_i$ and $w_\gth$ define the
points in the toric diagram corresponding to divisors of the
resolution.

Next, we describe how to associate to the toric diagram a toric variety
which represents the resolution of $\Cplx^n/G\,$. Each of the vectors,
$v_i$ and $w_\gth\,$, correspond to a homogeneous coordinate, $z_i$
and $x_\gth\,$, of the resolution $\Res(\Cplx^n/G)\,$,  respectively.
As in the previous section, the divisors are defined by setting the
corresponding coordinate to zero: 
\equ{
D_i ~=~  \big\{\, z_i ~=~0\, \big\}~, 
\qquad 
E_\gth ~=~  \big\{\, x_\gth ~=~0\, \big\}~.
\labl{defDivs}
}
The ordinary divisors $D_i$ are never compact, while the exceptional
divisors are compact only when the lie in the interior of the toric
diagram. Introduce a set of local coordinates
\equ{
Z_j ~=~ \prod_i z_i^{(v_i)_j} \, \prod_\gth x_\gth^{(w_\gth)_j}~,
\labl{localCnG}
}
where $(v_i)_j$ denotes the $j$th component of the vector $v_i\,$.
We can read off the $n$ linear equivalence relations of the divisors
from them:
\equ{
\sum_i (v_i)_j\, D_i \,+\, \sum_\gth (w_\gth)_j\, E_\gth ~\sim~ 0~.
\labl{lineqvCnG}
}
At the same time the $(\Cplx^*)^N$ group of scaling of homogeneous
coordinates $z_i$ and $x_\gth$ is defined, such that it leaves the
local coordinates~\eqref{localCnG} invariant. This means, that if one
substitutes the scaling charges as values of the divisors in the
linear equivalence relations~\eqref{lineqvCnG} one obtains zero. 
The action $(\Cplx^*)^N$ of scaling is in general not well--defined on
$\Cplx^{n+N}\,$.  The resolution of the $\Cplx^n/G$ orbifold is
defined as  
\equ{
\Res(\Cplx^n/G) ~=~ \big( \Cplx^{n+N} \,-\, F \big)/ (\Cplx^*)^N~, 
}
where exclusion set $F$ is defined, as in the previous section, such
that in non of the coordinate patches singularities arise. This
coincides with the definition of the exclusion set given
in~\cite{Cox:1993fz}.

To obtain the integrals of the various divisors over the resolution,
loosely speaking the intersection numbers, assume that the definition
of the toric diagram has to be completed by giving a triangulation. In
this section we assume that the triangulation is 
unique. In section~\ref{sc:multipleresolutions} we return to the
complication when more than one triangulation is possible. 
The triangulation defines the basic cones, i.e.\ the smallest possible
cones, inside the toric diagram. The intersection of the divisors, or
the corresponding integral, that form the corners of the basic cones,
are defined to have unity intersection number. In other words, the
triangulation defines the compact curves of the resolution as the
interior lines in the toric diagram. The intersection number with the
divisor of the basic cone of which such a compact curve forms the edge
is equal to one. In addition, the intersection of divisors that are linearly
dependent vanishes. In the projected toric diagram in three complex
dimensions this corresponds to the case when three or more divisors
are aligned. 
The set of basic cones, together with the linear equivalence
relations, determine all other integrals of the divisors 
uniquely. In total there are: 
\equ{
\#_2(D,E)  ~=~ \frac {(N+2)(N+3)}{2}~,
\qquad 
\#_3(D,E) ~=~ \frac{(N+5)(N+4)(N+3)}{6}~, 
}
such integrals in two and three complex dimensions,
respectively. When there are a large number of exceptional divisors,
this means, that the total number of integrals grows
rapidly. Indeed, in three complex dimensions we have  
$\#_3(D,E) =  20,\, 35,\, 56,\, 84\,$, for $N = 1,2,3,4$ exceptional
divisors. (The resolution of the $\Intr_{6-II}$ singularity provides an
example of the case with $N=4\,$.) Fortunately, we do not need to give
all these integrals explicitly, because we can use the
linear equivalences to express integrals involving ordinary
divisors in terms of those involving exceptional divisors only. The
number of integrals of exceptional divisors in two and three complex
dimensions, grows like 
\equ{
\#_2(E) ~=~ \frac{N(N+1)}{2}~,
\qquad 
\#_3(E) ~=~ \frac {(N+2)(N+1)N}{6}~, 
}
with the number of exceptional divisors $N\,$. In particular, in three
complex dimensions we find the more manageable numbers  
$\#_3(E) = 1,\, 4,\, 10,\, 20\,$ for $N=1,2,3,4\,$. This
completes the purely geometrical description of resolutions of
$\Cplx^n/G$ singularities.

For applications in model building of heterotic orbifold blowups we
need to specify the gauge background. The simplest gauge backgrounds,
apart from the standard embedding,  are $\U{1}$ line bundle
backgrounds. As we have seen above, complex line bundles play a
prominent role in the toric geometrical description of orbifold
resolution. Taking the linear equivalence relations~\eqref{lineqvCnG}
into account, a basis for $\U{1}$ gauge backgrounds is given by the
exceptional divisors. In general they do not represent the minimal
line bundles of the resolution. A basis of the smallest line bundles
is obtained by requiring, that each of the element integrated on all 
compact curves, that form a basis for all compact curves, either gives
zero or one. In the two 
dimensional case all exceptional divisors are compact. In three 
complex dimensions all curves, represented by lines between two
adjacent divisors, that go through the interior of the toric diagram,
are compact. Taking into account the linear equivalences again, one
can define such a basis of $N$ minimal compact curves $C_\gth$ of the
resolution. After that it is a straightforward exercise in linear
algebra to find those linear combinations $\go_\gth$ of exceptional
divisors, that are orthonormal to the basis of compact curves
\equ{
\int_{C_\gth} \go_{\gth'} ~=~ \gd_{\gth, \gth'}~. 
}
This basis of $N$ compact curves can be used to compute the weights of
the $N$ scalings defining the $(\Cplx^*)^N\,$. To find the relevant
charges, one may compute the intersections between these compact
curves and all divisors.

After this basis has been determined, the general $\U{1}$ gauge bundle
embedded in the $\SO{32}$ or $\E{8}\times \E{8}'$ gauge group, can be
represented as 
\equ{
\frac{\cF_V}{2\gp} ~=~ 
\sum_{[\gth]} V_\gth^I\, \go_\gth\, H_I~. 
}
For each representative $[\gth]$ the vector $V_\gth$ either contains
only integers or only half 
integers. This ensures, that we have well--defined eigenvalues on the
roots of the adjoint of $\SO{32}$ super Yang--Mills theory. (When we
want to discuss compactification of $\E{8}\times \E{8}$ SYM or either
heterotic string, we need that the entries of $V_\gth$ sum to an
even number.) In analogy to~\eqref{IdOrbiShiftCnZn} we can make
identifications of the vectors $V_\gth$ and the orbifold gauge shift
vectors $v_i$ for each of the Abelian factors inside the orbifold 
group $G\,$. The integral of $\cF_V$ over each non-compact divisor
$D_i$ gives rises to such an relation. This procedure does not work
when on a face of the toric diagram, one or more exceptional divisors
are located. In such a case, the face defines the resolution of a
suborbifold $\Cplx^2/G'\,$, $G' \subset G\,$. To make the
identification of the orbifold and line bundle shifts, one has to
perform the matching on this subvariety. To restrict the divisors to
this resolution of the suborbifold, one needs to put some exceptional
divisors to zero. This mean ignoring the corresponding extra
homogeneous coordinate and its associated $\Cplx^*$ scaling. In this
way all properties, including e.g.\ the total Chern class, can be
reduced to the subresolution.

Only those gauge configurations which in addition satisfy the
integrated Bianchi identity  
\equ{
\int_{C_2} \tr \cR^2 ~=~ \int_{C_2} \tr \cF_V^2~, 
}
for all compact 2--cycles $C_2$, define consistent background on the
resolution. In this work we will often require, that the integrated
Bianchi identity also vanishes for non--compact 2--cycles. The latter
requirement is not necessary, but we will see in examples, that with
this condition we are able to recover many of the modular invariant
heterotic orbifold models. In particular, for $\Res(\Cplx^2/G')\,$, the
resolution is itself the only 2--cycle, which obviously is
non--compact. For the three dimensional case, the (non--)compact
holomorphic 2--cycles correspond to the (non--)compact divisors.

As a final cross check on the validity of the application of toric
methods to obtain resolutions of heterotic orbifold, we compute the
four dimensional spectra. We only compute the spectra of those models,
that satisfy all possible consistency conditions. (For the other
models, there is $H$ flux flowing out of the singularity, this means
that the resolution has locally torsion. Therefore the standard index
theorems for computing the spectra do not apply.) The four dimensional
spectrum on the resolution with the $\U{1}$ gauge background can be
computed using the multiplicity operator 
\equ{
N_V ~=~ \int \Big\{\,  
\frac 1{3!} \, \big(\frac {\cF_V}{2\gp} \big)^3 \,+\,  
\frac 1{12}\,c_2(\cR)\, \frac {\cF_V}{2\gp} 
\,\Big\}~. 
\labl{multiplicity}
}
This operator can then be applied to the branching of the adjoint
representation due to the gauge background to determine the
multiplicity factors. As we are considering resolutions of
non--compact orbifolds, the multiplicities often take fractional
values.

After this general digression of the use of toric geometrical techniques
to obtain resolutions of heterotic orbifold models, we give in the
following two subsections interesting examples of orbifold resolutions,
$\Res(\Cplx^2/\Intr_3)$ and $\Res(\Cplx^3/\Intr_4)\,$, which both have
two exceptional divisors.

\subsection{Resolution of $\boldsymbol{\Cplx^2/\Intr_3}$} 
\labl{sc:C2Z3}

\begin{figure}
\begin{center} 
\begin{tabular}{ c}
\raisebox{0ex}{\scalebox{0.7}{\mbox{\input{C2Z3.pstex_t}}}}
\end{tabular} 
\end{center}
\captn{\label{fg:toricC2Z3}
The toric diagram of $\Res(\Cplx^2/\Intr_3)$ is displayed. Both
exceptional divisors $E_1$ and $E_2$ are compact. 
}
\end{figure}

To illustrate the resolutions with more than one exceptional divisor
in two dimensions, we consider the resolution of $\Cplx^2/\Intr_3\,$,
as an example. The orbifold action reads
\equ{
\gth\,:~ \big( \tZ_1,  \tZ_2\big) ~\ra~ 
\big( e^{2\gp i \gf_1} \tZ_1, 
e^{2\gp i \gf_2} \tZ_2 \big)~, 
\qquad 
\gf ~=~ \frac 13 \big( 1,\, 2\big)~. 
}
Taking the vectors, $v_1=(1,0)$ and $v_2=(0,1)\,$, to represent the
ordinary divisors, $D_1$ and $D_2\,$, in the toric diagram, we find,
that, $w_1 = \frac 13( 1,2)$ and $w_2 = \frac 13(2,1)\,$, indicate the
two exceptional divisors $E_1$ and $E_2\,$, respectively. The
resulting toric diagram of the resolution is given in
figure~\ref{fg:toricC2Z3}. From the local coordinates~\eqref{localCnG} 
\equ{
Z_1 ~=~ z_1 \, x_1^{\frac 13}\, x_2^{\frac 23}~, 
\qquad 
Z_2 ~=~ z_2 \, x_1^{\frac 23}\, x_2^{\frac 13}~, 
}
we read off the linear equivalence relations 
\equ{
3\, D_1 \,+\, E_1 \,+\, 2\, E_2 ~\sim~ 0~, 
\qquad 
3\, D_2 \,+\, 2\, E_1 \,+\, E_2 ~\sim~ 0~, 
}
and the $(\Cplx^*)^2$ scalings 
\equ{
\big(z_1,\, z_2,\, x_1,\, x_2 \big) ~\sim~
\big(\gl_1\inv \,z_1,\, \gl_2\inv \,z_2,\, 
\gl_2^2\gl_1\inv \,x_1,\, \gl_1^2 \gl_2\inv \,x_2 \big)~. 
}
The exclusion set reads 
\equ{
F ~=~ \{\, z_1 = x_1 = 0 \,\} 
~\cup~
\{\, z_2 = x_2 = 0 \,\}
~\cup~ 
 \{\,  z_1 = z_2 = 0 \, \}~, 
}
as can be seen from the toric diagram displayed in
figure~\ref{fg:toricC2Z3}.

\intersectionsCtwoZthree

From this toric diagram one can read off the basic cones: 
\equ{
D_1 E_2 ~=~ E_1 E_2 ~=~ D_2 E_1 ~=~ 1~.
}
Because the toric variety is two complex dimensional the divisors are
the same as the curves of the resolution, all intersection of curves with
divisors can be compactly displayed in a single table, see
table~\ref{tb:intersectionsC2Z3}. From the intersection table we infer
that, $D_2$ and $D_1\,$, define $(1,1)$--forms that are orthonormal to the
compact curves, $E_1$ and $E_2\,$, respectively. Hence we can expand a
$\U{1}$ gauge background as 
\equ{
\frac {\cF_{V}}{2\gp} ~=~ 
\big(\, V_1^I\, D_1 \,+\, V_2^I\, D_2 \big)\, H_I~, 
\labl{backgroundC2Z3}
}
where, $V_1$ and $V_2\,$, are either integer or half integer vectors. 
Using methods explained above, we can make identifications between
the orbifold gauge shift, $v\,$, and the vectors, $V_1$ and $V_2\,$, by
computing the integrals over $\cF_V$ over non--compact curves, $D_1$
and $D_2\,$, respectively: 
\equ{
v^I H_I ~\equiv~ \int_{D_1} \frac {\cF_{V}}{2\gp} ~=~  
-\frac 13\big(2\, V^I_1 \,+\, V^I_2\big) H_I~, 
\quad 
- v^I H_I ~\equiv~ \int_{D_2} \frac {\cF_{V}}{2\gp} ~=~  
-\frac 13\big( V^I_1 \,+\, 2\, V^I_2 \big) H_I~. 
}
It follows that $V_1 \equiv - V_2 \equiv 3v\,$, in order that the
line bundle background can be interpreted in the blow down limit.

\modelsCtwoZthree

To determine the consequences of the Bianchi identity, we compute the
integral of the second Chern class over the resolution 
\equ{
-\frac 12\, \int \frac{\tr\, \cR^2}{(2\gp i)^2} ~=~ \int c_2(\cR) ~=~ \frac 83~.
}
Requiring that the integrated Bianchi identity vanishes, leads to the
consistency condition
\equ{
V_1^2 \,+\, V_2^2 \,+\, V_1\cdot V_2 ~=~ 8~. 
}
This is the analog of the modular invariance consistency condition of
the heterotic string
\equ{
(3v)^2 ~=~ 2 ~\text{mod}~6.  
}
In table~\ref{tb:modelsC2Z3} we give the inequivalent modular
invariant orbifold gauge shifts, $v\,$, and indicate the vectors,
$V_1$ and $V_2\,$, of the corresponding blowup model(s). The first
four orbifold models in this table can be realized in blowup with the
choice:  $V_2 = -V_1\,$. For the orbifold standard embedding 
$3v =(1^2, 0^{14})$ can also be realized in an alternative way, in
which the vectors are not simply equal and opposite, but
nevertheless satisfy  the condition that they can be identified with
the orbifold gauge shift.

The final orbifold model in table~\ref{tb:modelsC2Z3} can not be
realized by any combination of resolution vectors, $V_1$ and $V_2\,$,
satisfying all conditions. For this reason we have separated it from
the rest of the table. We give two proposals of vectors that could
realize the blowup of the orbifold model: The first realization has
vectors, $V_1$ and $V_2\,$, that can be directly identified with the
orbifold one, but do not have a vanishing Bianchi identity. 
The second realization has vectors $V_1$ and $V_2\,$, that lead to
the vanishing of the Bianchi identity, but cannot be linked directly
to an orbifold shift. For this model and all the others where we can
compute the spectrum, they coincide with the ones that were 
identified in~\cite{Honecker:2006qz}.

\subsection{Resolution of $\boldsymbol{\Cplx^3/\Intr_4}$}
\labl{sc:C3Z4}

The second example of a resolution with two exceptional divisors is
obtained from the three dimensional orbifold $\Cplx^3/\Intr_4\,$: 
\equ{
\gth\,:~ \big( \tZ_1,  \tZ_2, \tZ_3 \big) ~\ra~ 
\big( e^{2\gp i \gf_1} \tZ_1, 
e^{2\gp i \gf_2} \tZ_2, 
e^{2\gp i \gf_3} \tZ_3 \big)~, 
\qquad 
\gf = \frac 14 \big( 1,\, 1,\, 2\big)~. 
}
The elements $\gth$ and $\gth^3$ are each other's complex conjugates,
hence there are two exceptional divisors $E_1$ and $E_2\,$. The
vectors 
\equ{
w_1 ~=~ \frac 14\, v_1 \,+\, \frac 14\, v_2 \,+\, \frac 12\, v_3~, 
\qquad 
w_2 ~=~ \frac 12\, v_1 \,+\, \frac 12\, v_2~, 
}
take the form, $\frac 14(1,1,2)$ and $\frac 12(1,1,0)\,$, in the
basis, $v_1 = (1,0,0)\,$, $v_2 = (0,1,0)\,$, $v_3 = (0,0,1)\,$,
respectively. This leads to the local coordinates 
\equ{
Z_1 ~=~ z_1 \, x_1^{\frac 14}\, x_2^{\frac 12}~,
\qquad
Z_2 ~=~ z_2 \, x_1^{\frac 14}\, x_2^{\frac 12}~,
\qquad
Z_3 ~=~ z_3 \, x_1^{\frac 12}~, 
}
which imply the linear equivalence relations
\equ{
4\, D_1 \,+\, E_1 \,+\, 2\, E_2 ~\sim~ 0~, 
\quad 
4\, D_2 \,+\, E_1 \,+\, 2\, E_2 ~\sim~ 0~, 
\quad 
2\, D_3 \,+\, E_1 ~\sim~ 0~. 
\labl{lineqvC3Z4}
}
The $(\Cplx^*)^2$ scalings 
\equ{
\big(z_1,\, z_2,\, z_3,\, x_1,\, x_2\big) ~\sim~ 
\big(\gl_1\inv \,z_1,\, \gl_1\inv \,z_2,\,  \gl_3\inv \,z_3,\, 
\gl_3^2 \, x_1,\, \gl_1^2 \gl_3\inv \, x_2 \big)~, 
\labl{scalingC3Z4}
}
require that the exclusion set is given by
\equ{
F ~=~ \{\, z_1 = z_2 = 0 \,\} 
~\cup~
\{\, z_3 = x_2 = 0 \,\}~, 
\labl{exclusionC3Z4}
}
in order to avoid singularities in any of the coordinate patches. 
The projected toric diagram was composed using the basis, 
$v_1=(0,0,1)\,$, $v_2=(1,0,1)\,$,  $v_=(0,1,1)\,$, in which  
$w_1=(\frac 14, \frac 12, 1)$ and $w_2=(\frac 12, 0, 1)\,$.

\begin{figure}
\begin{center} 
\begin{tabular}{ c }
\raisebox{0ex}{\scalebox{0.5}{\mbox{\input{C3Z4.pstex_t}}}}
\end{tabular} 
\end{center}
\captn{\label{fg:toricC3Z4}
This figure gives the projected toric diagram of
$\Res(\Cplx^3/\Intr_4)\,$. Only the exceptional divisor $E_1$ is
compact, all other divisors are non--compact. 
}
\end{figure}

The projected toric diagram~\ref{fg:toricC3Z4} implies that the basic
cones  
\equ{
D_1\, E_1\, E_2 ~=~ D_2\, E_1\, E_2 ~=~ 
D_1\, D_3\, E_1 ~=~ D_2\, D_3\, E_1 ~=~ 1~, 
}
all have unit intersection number, and that the integrals 
\equ{
D_1\, D_2\, E_2 ~=~ D_3\, E_1\, E_2 ~=~ 0
}
vanish. The total number of integrals of divisors on this resolution is
$35$, but as discussed above it suffices to only give the $4$
integrals of the exceptional divisors
\equ{
E_1^2\, E_2 ~=~0~, 
\quad 
E_2^2\, E_1 ~=~ -2~, 
\quad 
E_1^3 ~=~ 8~, 
\quad 
E_2^3 ~=~ 2~, 
}
as all other integrals can be determined from them using the
linear equivalences~\eqref{lineqvC3Z4}.

The exceptional divisor $E_1$ lies in the interior of the projected
toric diagram, and hence is compact. This can be easily confirmed
explicitly. The divisor $E_1$ is embedded as 
\equ{
E_1 ~=~ 
\big(\gl_1\inv \,z_1,\, \gl_1\inv \,z_2,\,  \gl_3\inv \,z_3,\, 
0,\, \gl_1^2 \gl_3\inv \, x_2 \big)~, 
}
inside the toric variety $\Res(\Cplx^3/\Intr_4)\,$. By fixing the
scaling such that $|\gl_1|^2 = |z_1|^2 + |z_2|^2$ and 
$|\gl_2|^2 = |z_3|^2 + |\gl_1^2 x_2|^2\,$, it is obvious that $E_1$ is
bounded and hence compact. Moreover, notice the coordinates $z_1$ and
$z_2$ have a scaling factor $\gl_1\inv$ and the coordinates $z_3$ and
$x_2$ have a scaling factor $\gl_3\inv$. Ignoring the factor
$\gl_1^2\,$, that also scales $x_2\,$, $E_1$ would be a direct product
of two $\CP^1$'s. However, precisely this additional scaling
of $x_2$ with $\gl_1^2$ means, that $E_1$ is not simply a direct
product of two $\CP^1$'s, but rather an $\CP^1$ bundle over
$\CP^1\,$. Such a surface is called the Hirzebruch surface
$\mathbb{F}_2\,$ in the mathematical literature.

The exceptional divisor $E_2$ is non--compact in three complex
dimension. It equals a direct product $\CP^1 \times \Cplx\,$, which
signals that we should view the situation from a two dimensional
complex perspective instead. The
edge of the toric diagram, in figure~\ref{fg:toricC3Z4}, spanned by
$D_1$ and $D_2\,$, is itself precisely the toric diagram of the
resolution $\Res(\Cplx^2/\Intr_2)\,$, as depicted on the left of
figure~\ref{fg:toricCnZn}. Therefore, the integrals computed in
subsection~\ref{sc:toricCnZn}, for $n=2\,$, can be directly applied to
the divisors, $D_1\,$, $E_2$ and $D_2\,$. Hence, in particular, we
have $D_1 E_2 = D_2 E_2 = 1\,$.

\intersectionsCthreeZfour

Next, we want to find a basis of orthonormal $(1,1)$--forms, that 
can used to expand the $\U{1}$ gauge background around. To determine 
this basis, we note that there exist four compact curves: 
$D_1 E_1\,$, $D_2 E_1\,$, $D_3 E_1\,$, and $E_1 E_2\,$.
Using the linear
equivalences~\eqref{lineqvC3Z4} we infer, that if we have constructed 
an orthonormal basis of $(1,1)$--forms on the curves, $D_1 E_1$ and
$E_1 E_2\,$, they are integer on all these compact curves. Such a
basis of $(1,1)$--forms is spanned by $D_1$ and $D_3\,$, see  
the same table~\ref{tb:intersectionsC3Z4}. This means that we can
expand the gauge background as  
\equ{
\frac {\cF_V}{2\gp} ~=~ -\frac 12\, E_1\, H_1 
\,-\, \frac 14\, (E_1 + 2 \, E_2) H_2~, 
\labl{backgroundC3Z4}
}
where $H_1 = V_1^I H_I$ and $H_2 = V_2^I H_I$, respectively. 
We have used the linear equivalences~\eqref{lineqvC3Z4} to
express $D_1$ and $D_3$ in terms of the exceptional divisors only.

In order that this gauge background~\eqref{backgroundC3Z4} defines a 
consistent compactification, we have to require that the Bianchi
identity vanishes when integrated over the compact divisor $E_1\,$. 
To determine the resulting condition we evaluate the second Chern
class 
\equ{
c_2(\cR) ~=~ D_1^2 \,-\, 2 \,D_1 D_3 \,-\, 2\, D_3^2 \,+\, 2\, D_1 E_2 
\,-\, D_3 E_2~, 
}
which leads to the necessary consistency condition
\equ{
V_1^2 \,+\, V_1\cdot V_2 ~=~ 4~. 
\labl{NESbianchiC3Z4}
}
This condition ensures, that the gauge background, defined by $V_1$
and $V_2\,$, is consistent.

\modelsCthreeZfour

In addition to this necessary condition, we may also require that
the integrated Bianchi vanishes on $E_2\,$, and on the subvariety
$\Res(\Cplx^2/\Intr_2)\,$. As noted above, the edge of the toric
diagram, figure~\ref{fg:toricC3Z4}, spanned by $D_1$ and $D_2\,$,
defines the toric diagram of $\Res(\Cplx^2/\Intr_2)\,$. This tell us,
that we should do the computation on two complex dimensional toric
variety, with the divisors $D_1$, $D_2$ and the exceptional one
$E_2\,$. All properties of this subvariety are inherent from
$\Res(\Cplx^3/\Intr_4)$ by setting $E_1=0\,$, i.e.\ simply ignoring
the homogeneous coordinate $x_1$ and its associated scaling
$\gl_3\,$. Indeed, the scaling~\eqref{scalingC3Z4} reduces to
\equ{
\big(z_1,\, z_2,\, z_3,\, x_2\big) ~\sim~ 
\big(\gl_1\inv \,z_1,\, \gl_1\inv \,z_2,\, z_3,\, 
 \gl_1^2 \, x_2 \big)~, 
}
which defines the space $\Res(\Cplx^2/\Intr_2) \times \Cplx\,$. 
It is also not difficult to check, that the total Chern class of
$\Res(\Cplx^3/\Intr_4)$ with vanishing $E_1$ reduces to that of
$\Res(\Cplx^2/\Intr_2)\,$.  Similarly, taking $E_1 = 0$
in~\eqref{backgroundC3Z4} gives us the gauge background on this
subresolution. This gives rise to the additional
conditions 
\equ{
V_1\cdot V_2 ~=~ -2~, 
\qand 
V_2^2 ~=~ 6~, 
\label{EXTRAbianchiC3Z4}
}
respectively.

Finally, we can make a partial matching with the orbifold gauge
shift. From the six dimensional perspective we can use
the identification of the orbifold and blowup shifts on the 
subresolution of $\Cplx^2/\Intr_2\,$. By integrating the bundle
background  over $D_1$ within  $\Res(\Cplx^2/\Intr_2)$ gives  
\equ{\labl{identifyZ4shiftbis}
2\, v^I\, H_I ~\equiv~ 
\int_{D_1} \frac {\cF_V}{2\gp} ~=~ -\frac 12 V_2^I\, H_I~.
}
We can identify this integral with the $\Intr_2$ gauge orbifold
shift $2\, v\,$. The identification from the four dimensional
perspective is more complicated, and will not be discussed here.

We can give a complete classification of all consistent models on the
resolution of $\Cplx^3/\Intr_4\,$, using {\em all} the conditions
described above. Table~\ref{tb:modelsC3Z4} gives the gauge shift
vectors of the possible heterotic orbifold models, and the vectors
$V_1$ and $V_2\,$, that define the $\U{1}$ bundle background on the
resolution. Only for the orbifold model numbered 4 in
table~\ref{tb:modelsC3Z4} we have not found a blowup model.This
orbifold model has no matter in the first twisted sector. Since the
blowup modes are precisely the twisted states of the string, we expect
that no complete resolution of this orbifold model exists.

For each of the other models, we compute the spectrum
using~\eqref{multiplicity}, and compare it with  the spectrum of the
corresponding orbifold model. The multiplicity operator takes the form  
\equ{
N_V ~=~ \frac 1{6} \Big[\, 
\frac 32 \big( \frac 12 \,-\, H_1^2 \big) H_2 \,+\, 
\big( 1 \,-\, H_1^2 \big) H_1
\,\Big]~, 
}
where we employed the short hand notation $H_i = V^I_i \, H_I\,$. 
The resulting spectra in the $\SO{32}$ theory are given in
tables~\ref{tb:spectraC3Z4} and~\ref{tb:spectraC3Z4bis}. The
multiplicity factors of $1/8$ and $1/4$ can be easily understood from
the heterotic orbifold point of view: In paper~\cite{Nibbelink:2003rc}
the local anomalies at four and six dimensional fixed points of
$T^6/\Intr_4$ were computed, using general trace formulae on
orbifolds~\cite{GrootNibbelink:2003gd}: The ten dimensional states
contribute $1/8$ of an anomaly at a $\Intr_4$ fixed point, the six
dimensional second--twisted sector contributes $1/4\,$,  and the four
dimension single--twisted sector gives integral contributions. The matter
representations can also be traced back to the orbifold model. The six
and four dimensional spectra of the heterotic string on $\Cplx^3/\Intr_4$ 
can be found in~\cite{Choi:2004wn,Nilles:2006np}. The spectra in  
tables~\ref{tb:spectraC3Z4} and~\ref{tb:spectraC3Z4bis} are obtained
from simple branching w.r.t.\ the unbroken gauge group, upto possible
mismatches due to vector--like states. Mostly only a single scalar is
not part of the charged chiral  spectrum on the resolution (as explained
in~\cite{GrootNibbelink:2007ew} this state has become a model
dependent axion part of the expansion of $B_2$).  
Some model have $\SU{N}$ gauge groups and therefore,
non--Abelian gauge anomalies could arise. However, from
tables~\ref{tb:spectraC3Z4} and~\ref{tb:spectraC3Z4bis} it can be
confirmed, that all pure $\SU{N}\,$, $N \geq 3\,$, anomalies
cancel. The models contain a bunch of $\U{1}$'s, that are all
potentially anomalous, we expect that their anomalies are canceled via
the Green--Schwarz mechanism involving universal and non--universal 
axions~\cite{Atick:1987gy,Dine:1987gj,Blumenhagen:2005pm,GrootNibbelink:2007ew}.

\section{Orbifolds with multiple resolutions}
\labl{sc:multipleresolutions}

\subsection{Generalities of multiple triangulations}
\labl{sc:multitriangulations}

In the general discussion and in the examples so far we have avoided
one further complication of generic resolutions of orbifold
singularities in three (or more) complex dimensions: The resolution of
a given $\Cplx^3/G$ orbifold might be non--unique. This difficulty 
arises precisely when more than one triangulation of the toric
diagram is possible. For clarity we first indicate which properties of
orbifold resolutions described and illustrated in
section~\ref{sc:multipleexceptionaldivisors} still hold, and after
that focus on novelties, that arise from the possibility of having
multiple triangulations.

Essentially all the properties of a resolution, discussed in
subsection~\ref{sc:generalitiesOrbi}, that do not depend on the
triangulation of the toric diagram, can be extended to orbifolds
which have non--unique resolutions. In particular, the
definition of the (exceptional) divisors~\eqref{defDivs}, the
construction of a set of local coordinates~\eqref{localCnG}, the linear
equivalences~\eqref{lineqvCnG}, and the $(\Cplx^*)^N$ scaling,  are
uniquely defined for any triangulation. As we have seen resolutions of
three dimensional orbifolds may contain two dimensional resolutions as
subvarieties. These subvarieties are identified as the faces of the 
toric diagram. Even though the resolution of three dimensional
orbifolds may not be unique, the toric diagrams corresponding to the
faces is uniquely defined by the divisors on them. Hence these
subvarieties are the same for each resolution.

The exclusion set $F$ does depend on the
triangulation~\cite{Cox:1993fz}: As before, the exclusion is defined
such that the resolution is by definition non--singular. In addition,
the curves, that are not realized as lines within the triangulation,
are part of the exclusion set. The latter makes the exclusion set
dependent on the triangulation of the toric diagram.

The integrals of the divisors over the resolution also crucially depend
on the triangulation: As described in subsection~\ref{sc:generalitiesOrbi}
the triangulation identifies the compact curves have unit intersection
number with some divisors of  the resolution. Hence, if the
triangulation is not unique, one can assign different intersection of
the compact curves with the divisors. The problem is, that there are
more basic cones possible in the toric diagram given the divisors
only, than can be realized in a given triangulation. This issue is
illustrated by the toric diagrams of the resolution of
$\Cplx^3/\Intr_2\times\Intr_2'\,$: Of the ten possible basis cones,
only four are realized within a triangulation, as we discuss in detail
in subsection~\ref{sc:C3Z22}. To define the integrals of the divisors,
interpreted as $(1,1)$--forms, over the resolution,  we employ the
following rules for any given triangulation:  
\enums{
\item The basic cones, that do exist within the triangulation, 
are formed by divisors with unity intersection number;
\item while those, that do not exist within the triangulation,
have intersection number zero. 
\item Any set of three divisors aligned in the projected toric
diagram, have vanishing integral. 
\item All other integrals of three divisors are obtained from these
defining ones, using linear equivalence relations.  
}
The first three rules give consistent assignments that do not clash
with the linear equivalence relations. Even though, these rules 
might in general be insufficient to determine all integrals of the
exceptional divisors, they are sufficient for the resolutions
considered in this paper. 
As in the previous sections, it may happen that the integral over some
divisors is non--vanishing due to the linear equivalence relations,
even though, as hypersurface the intersection of these divisors is
excluded. As we will show in the examples of resolutions of
$\Cplx^3/\Intr_2\times \Intr_2'$ below, using the definition
of the integral of divisors given here, we are able to obtain blowup
versions of all heterotic models on this orbifold. In addition, we
obtain their spectra, which are all free of non--Abelian anomalies.

\subsection{Resolutions of
$\boldsymbol{\Cplx^3/\Intr_2\times\Intr_2'}$}
\labl{sc:C3Z22}

\begin{figure}
\begin{center} 
\begin{tabular}{ c c c }
\raisebox{0ex}{\scalebox{0.5}{\mbox{\input{C3Z22sym.pstex_t}}}}
& \qquad \qquad & 
\raisebox{0ex}{\scalebox{0.5}{\mbox{\input{C3Z22E1.pstex_t}}}}
\end{tabular} 
\end{center}
\captn{\label{fg:toricC3Z22}
The two inequivalent (projected) toric diagram of 
$\Res(\Cplx^3/\Intr_2\times\Intr_2')$ are displayed. The left one we 
call the ``symmetric'' resolution, while the right one the ``$E_1$''
resolution. 
}
\end{figure}

We consider $\Cplx^3/\Intr_2\times\Intr_2'$ as an
example of an orbifold, that admits more than one resolution. To
clearly separate which statements are triangulation dependent, and
which are not, we first describe those properties that are valid for
each resolution. After  that we compute the integrals of the divisors
on the two inequivalent resolutions separately. Finally we, study the
relation between heterotic models on this orbifold, and its possible
resolutions.

\subsubsection*{Triangulation independent properties of the resolutions}

The orbifold $\Cplx^3/\Intr_2\times \Intr_2'$ is defined by the three
$\Intr_2$ orbifold actions: 
\equ{
\arry{llrcl}{\dsp 
\gth\,: &\big( \tZ_1,  \tZ_2, \tZ_3 \big) ~\ra~ 
\big( \tZ_1, -\tZ_2, -\tZ_3 \big)~, 
& \quad 
\gf &=& \frac 12 \big( 0, 1, 1 \big)~, 
\\[2ex] \dsp 
\gth'\,: & \big( \tZ_1,  \tZ_2, \tZ_3 \big) ~\ra~ 
\big( -\tZ_1, \tZ_2, -\tZ_3 \big)~, 
& \quad 
\gf' &=& \frac 12 \big( 1, 0, 1 \big)~, 
\\[2ex] \dsp 
\gth\gth'\,: & \big( \tZ_1,  \tZ_2, \tZ_3 \big) ~\ra~ 
\big( -\tZ_1, -\tZ_2, \tZ_3 \big)~, 
&\quad 
\gf+\gf' & =& \frac 12 \big( 1, 1, 0 \big)~, 
}  
}
where the latter can be viewed as the combination of the first two. This
orbifold has three twisted sectors, and hence three exceptional
divisors $E_1\,$, $E_2\,$, and $E_3\,$, defined by the vectors 
\equ{
w_1 ~=~ \frac 12\, v_2 \,+\, \frac 12\, v_3~,
\qquad 
w_2 ~=~ \frac 12\, v_1 \,+\, \frac 12\, v_3~,
\qquad 
w_3 ~=~ \frac 12\, v_1 \,+\, \frac 12\, v_2~.
}
In the standard basis for $v_i\,$, they lead to the local coordinates 
\equ{
Z_1 ~=~ z_1\, x_2^{\frac 12}\, x_3^{\frac 12}~, 
\qquad 
Z_2 ~=~ z_2\, x_1^{\frac 12}\, x_3^{\frac 12}~, 
\qquad 
Z_3 ~=~ z_3\, x_1^{\frac 12}\, x_2^{\frac 12}~, 
\labl{localC3Z22}
}
on the resolutions. This determines the linear equivalences 
\equ{
2\, D_1 \,+\, E_2 \,+\, E_3 ~\sim~ 
2\, D_2 \,+\, E_1 \,+\, E_3 ~\sim~ 
2\, D_3 \,+\, E_1 \,+\, E_2 ~\sim~ 0~. 
\labl{lineqvC3Z22}
}
Using these linear equivalences we can represent the second Chern
class as  
\equ{
c_2(\cR) ~=~ 
- \frac 34\, \big(\, E_1^2 \,+\, E_2^2 \,+\, E_3^2\,\big) 
\,-\, \frac 14\, \big(\, E_1\, E_2 \,+\,  E_2\, E_3 \,+\,  E_3\, E_1 \,\big)~. 
}

The $(\Cplx^*)^3$ action on the homogeneous coordinates can be 
parameterized as 
\equ{
\big(z_1,\, z_2,\, z_3,\, x_1,\, x_2\, x_3\big) ~\sim~ 
\big(\gl_2\inv\gl_3\inv \,z_1,\, \gl_1\inv\gl_3\inv \,z_2,\,  
\gl_1\inv\gl_2\inv \,z_3,\, 
\gl_1^2 \, x_1,\, \gl_2^2 x_2,\, \gl_3^2 \, x_3 \big)~.
\labl{scalingC3Z22}
}
The integrals 
\equ{
D_1\, D_2\, E_3 ~=~ 
D_2\, D_3\, E_1 ~=~ 
D_3\, D_1\, E_2 ~=~0 
} 
all vanish: they are aligned in the projected toric
diagram, see figure~\ref{fg:toricC3Z22}. But precisely these edges of
the projected toric diagrams define resolutions of 
$\Cplx^2/\Intr_2$ orbifolds, discussed in
section~\ref{sc:explicitblowups}. Hence each of these edges 
correspond to a six dimensional model. 
There are two inequivalent triangulations, which are displayed in
figure~\ref{fg:toricC3Z22}, which we now in turn describe.

\subsubsection*{The resolution with the ``symmetric'' triangulation}

We investigate the topological properties of the ``symmetric''
triangulation, defined on the left side of
figure~\ref{fg:toricC3Z22}. First of all, the exclusion set is defined
as  
\equ{
\arry{ll}{
F ~=\, &\dsp  \{\, z_1 = z_2 = 0 \,\} 
~\cup~
\{\, z_2 = z_3 = 0 \,\} 
~\cup~
\{\, z_1 = z_3 = 0 \,\}
\\[2ex] &\dsp    ~\cup~
\{\, z_1 = x_1 = 0 \,\}
~\cup~
\{\, z_2 = x_2 = 0 \,\}
~\cup~
\{\, z_3 = x_3 = 0 \,\}~. 
} 
\labl{exclusionC3Z22sym}
}
This ensures that there are no singularities, and that the dashed
lines in the left projected toric diagram in
figure~\ref{fg:toricC3Z22}, correspond to non--existing curves. 
We read off that the basic cones are given
by  
\equ{
D_1\, E_2\, E_3 ~=~ 
D_2\, E_3\, E_1 ~=~ 
D_3\, E_1\, E_2 ~=~
E_1\, E_2\, E_3 ~=~ 1~, 
\labl{basicConessym}
}
while the other possible basic cones, that are non--existent in this
triangulation, vanish: 
\equ{
\arry{c}{ \dsp 
D_1\, E_1\, E_2 ~=~ 
D_1\, E_1\, E_3 ~=~ 
D_2\, E_1\, E_2 ~=~ 0~, 
\\[2ex] \dsp 
D_2\, E_2\, E_3 ~=~ 
D_3\, E_1\, E_3 ~=~ 
D_3\, E_2\, E_3 ~=~ 0~. 
}
}
As we observed in section~\ref{sc:generalitiesOrbi} all 56 possible
integrals can be conveniently summarized by giving only the
10 involving the exceptional divisors only. Because of the high
amount of symmetry within the toric diagram, we can summarize all
integrals over the exceptional divisors as 
\equ{
E_p^3 ~=~ - E_p^2\, E_{q\neq p} ~=~ E_1\, E_2 \, E_3 ~=~ 1~. 
} 
From these integrals we easily compute the integrals over all
compact curves of all divisors. The result is tabulated in
table~\ref{tb:intersectionsC3Z22sym}.

\intersectionsCthreeZtwoSqsym

The curves that are not part of the triangulation do not exist in the
resolution as hypersurfaces. Nevertheless, we see in
table~\ref{tb:intersectionsC3Z22sym}, below the double line, that even
though curves, like $D_1 E_1$ do not exist, the integral 
$D_1 E_1 X\,$, of the dual $(2,2)$--form over $X$ ($X$ being $D_2\,$ 
or $D_3$ or $E_1\,$) does not  vanish.

\subsubsection*{The resolution with the ``$\boldsymbol{E_1}$'' triangulation}

Next we discuss the ``$E_1$'' triangulation. There are in fact two
other possible triangulations, ``$E_2$'' and ``$E_3$'',  but they are
simply obtained from this one by interchanging the labels $1$, $2$ and
$3$, hence do not constitute truly different resolutions. The
exclusion set reads in this case 
\equ{
\arry{ll}{
F ~=\, &\dsp  \{\, z_1 = z_2 = 0 \,\} 
~\cup~
\{\, z_2 = z_3 = 0 \,\} 
~\cup~
\{\, z_1 = z_3 = 0 \,\}
\\[2ex] &\dsp    ~\cup~
\{\, x_1 = x_2 = 0 \,\}
~\cup~
\{\, z_2 = x_2 = 0 \,\}
~\cup~
\{\, z_3 = x_3 = 0 \,\}~. 
} 
\labl{exclusionC3Z22E1}
}
All the basic cones of the``$E_1$'' triangulation contain the
exceptional divisor $E_1\,$: 
\equ{
D_1\, E_1\, E_2 ~=~ 
D_1\, E_1\, E_3 ~=~ 
D_2\, E_1\, E_3 ~=~
D_3\, E_1\, E_2 ~=~ 1~.  
}
In addition, we have the non--existing basic cones 
\equ{
\arry{c}{ \dsp 
D_1\, E_2\, E_3 ~=~ 
E_1\, E_2\, E_3 ~=~ 
D_2\, E_1\, E_2 ~=~ 0~, 
\\[2ex] \dsp 
D_2\, E_2\, E_3 ~=~ 
D_3\, E_1\, E_3 ~=~ 
D_3\, E_2\, E_3 ~=~ 0~. 
}
}
From this input data we obtain the following integrals of 
the exceptional divisors: 
\equ{
\arry{ccc}{\dsp 
E_1^2\, E_2 ~=~ E_1^2\, E_3 ~=~ 
E_2^2\, E_3 ~=~ E_3^2\, E_2 ~=~ 0~, 
&& \dsp 
E_1\, E_2\, E_3 ~=~ E_1^3 ~=~ 0~, 
\\[2ex] 
E_2^2\, E_1 ~=~ E_3^2\, E_1 ~=~ -2~, 
&& \dsp 
E_2^3 ~=~ E_3^3 ~=~ 2~. 
}
}

\intersectionsCthreeZtwoSqE

The integrals over the compact curves of the divisors can again be 
computed straightforwardly, using the linear equivalences. The
resulting integrals are listed in
table~\ref{tb:intersectionsC3Z22E1}. 
Also from this table we see, that setting all integrals that involve
$(2,2)$--forms dual to curves, that are not part of the triangulation
of the toric diagram, to zero, leads to inconsistencies. 
In this case only the curve $E_2 E_3$ has only
vanishing integrals, and hence is not in conflict with the linear
equivalence relations. Note, that also the divisor $E_1$ does not
intersect with any of the curves listed in
table~\ref{tb:intersectionsC3Z22E1}.

\subsection{Heterotic models from resolutions of
$\boldsymbol{\Cplx^3/\Intr_2\times \Intr_2'}$}
\labl{sc:hetC3Z22}

As described at the beginning of subsection~\ref{sc:C3Z22} many
topological properties are the same for all resolutions of
$\Cplx^3/\Intr_2\times \Intr_2'\,$. In particular, the six dimensional
analysis corresponding to the edges of the projected toric diagrams, 
figure~\ref{fg:toricC3Z22}, are independent on the resolution
chosen. Therefore, we begin with the resolution independent
properties in our construction of heterotic models on these
resolutions.

The gauge background on the resolution can in general be expanded as 
\equ{
\frac{\cF_V}{2\gp} ~=~ - \frac 12 
\big(\, H_1 \, E_1 \,+\, H_2 \, E_2 \,+\, H_3 \, E_3 \,\big)~,
}
where $H_1 = V_1^I H_I\,$, etc. 
To obtain the gauge configurations on the three edges of the projected
toric diagram, we only take the exceptional divisor into account which
lives on that particular edge. Using the analysis of
$\Res(\Cplx^2/\Intr_2)\,$, presented in section~\ref{sc:toricCnZn}, we
infer that $V_i$ have either only integer or half--integer
entries. In addition, we make the identification between the orbifold
gauge shift vectors $v_1\,$, $v_2$ and $v_3 \equiv v_1+v_2\,$. 
For example, on the edge spanned by $D_2$ and $D_3\,$, we
have 
\equ{
\int_{E_1} \frac{\cF_V}{2\gp} ~=~ V_1^I\, H_i~, 
\qquad 
v_1^I\, H_I ~\equiv~ 
\int_{D_2} \frac{\cF_V}{2\gp} ~=~ - \frac 12\, V_1^I\, H_i~.
}
The orbifold gauge shift vectors satisfy the modular invariance
conditions  
\equ{
(2 v_1)^2 ~=~ 2~\text{mod}~ 4~, 
\qquad 
(2 v_2)^2 ~=~ 2~\text{mod}~ 4~, 
\qquad 
(2 v_3)^2 ~=~ 2~\text{mod}~ 4~. 
}
Similarly, we know from the discussion in section~\ref{sc:toricCnZn},
that the integrated Bianchi identities on the three edges do not
necessarily have to vanish, but if they do, we find the conditions
\equ{
V_1^2~=~ V_2^2 ~=~ V_3^2 ~=~ 6~. 
\labl{6DBianchiC3Z22}
}

\subsubsection*{Heterotic model building on the ``symmetric''
resolution}

\modelsCthreeZtwoSqsym

We turn to the specific properties of the heterotic model
construction on the symmetric resolution. First of all we check the
quantization conditions 
\equ{
\int_{E_1E_2} \frac{\cF_V}{2\gp} ~=~ -\frac 12
\big( - V_1^I -V_2^I + V_3^I \big) H_I~, 
\label{quantfluxz2z2}
}
and cyclic permutation of the labels $1\,$, $2$ and $3\,$. 
The factor $1/2$ might seem worrying, but is in fact harmless, because
we know that in order to have an orbifold interpretation, we need that
$\frac{1}{2}V_3 = \frac{1}{2}(V_1 + V_2)$ modulo a vector in the adjoint
or in the spinorial representation of $\SO{32}$, and in both cases the
Dirac quantization condition \eqref{quantfluxz2z2} is satisfied.
The integrated Bianchi identities on the divisors
$E_1$, $E_2$ and $E_3$, give rise to the requirements: 
\equ{
V_1^2 \,+\, 2\, V_2 \cdot V_3 ~=~ 
V_2^2 \,+\, 2\, V_1 \cdot V_3 ~=~ 
V_3^2 \,+\, 2\, V_1 \cdot V_2 ~=~ 8~. 
\labl{4DBianchiC3Z22}
}
When combining this with the six dimensional Bianchi requirements, we
conclude that 
\equ{
V_1 \cdot V_2 ~=~ V_2 \cdot V_3 ~=~ V_1 \cdot V_3 ~=~ 1~.
}

The solution of these conditions and the corresponding orbifold models
are listed in table~\ref{tb:modelsC3Z22sym}. It is remarkable that the
orbifold shift vectors $2v_i$ and the vectors $V_i$ characterizing the
gauge bundle are almost identical. Indeed, a sign flip in some entries

of an orbifold shift is irrelevant, as well as the addition of vectors
in the lattice of the adjoint or the spinorial representations of $\SO{32}$. 
The four dimensional chiral spectrum on this resolution of the
$\Cplx^3/\Intr_2\times \Intr_2'$ can be computed from the multiplicity
operator 
\equ{
N_V = \frac 1{6} \big( H_1 + H_2 + H_3\big)
\Big[ 
\frac 12 \big( H_1 H_2 + H_2 H_3 +  H_3 H_1 \big)
- \frac 18 \big( H_1^2 + H_2^2 +  H_3^2 \big) 
- \frac 14 \Big]-\frac{3}{8}\,H_1 H_2 H_3~. 
\labl{multiplicityC3Z22sym}
}
The resulting spectra are rather elaborate, because of multiple
branchings, and not very illuminating, we refrain from giving them
explicitly in the paper. However, by direct inspection of these
spectra we confirmed that  all the models listed in
table~\ref{tb:modelsC3Z22sym} are free of irreducible anomalies.

\subsubsection*{Heterotic model building on the ``$\boldsymbol{E_1}$''
resolution}

For the other resolution, the quantization requires that: 
easily:
\equ{
\int_{E_1E_2} \frac{\cF_V}{2\gp} ~=~ V_2^I\, H_I~, 
\quad 
\int_{D_1E_1} \frac{\cF_V}{2\gp} ~=~ - \frac 12 (H_2+ H_3)~, 
\quad 
\int_{E_1E_3} \frac{\cF_V}{2\gp} ~=~ V_3^I\, H_I~. 
}
The quantization condition can only be satisfied only if $\frac{1}{2}(V_2+V_3)$
is a vector containing either only {\it even}, or only {\it odd} numbers.
Moreover, in order to have an identification with the orbifold models, we need
$\frac{1}{2} V_1=\frac{1}{2}(V_2+V_3)$ upto lattice vectors of the
adjoint or spinorial representation of $\SO{32}\,$. This implies that
$V_1$ contains either only odd, or only even numbers. When all entries
are odd $V_1^2 \geq 16\,$, while in the even case $V_1^2$ is a
multiple of four. In either case the Bianchi identity $V_1^2=6$ cannot
be satisfied. Thus, no model can be build in such a
resolution of the $\Cplx^3/\Intr_2\times \Intr_2'$ orbifold singularity,
that fulfils all the consistency conditions listed above.

\section{Conclusions}
\labl{sc:concl}

We have investigated resolutions of heterotic orbifolds using toric
geometry. Our initial motivation was to understand the topology behind
the recently constructed heterotic models on explicit blowup of
$\Cplx^n/\Intr_n$ singularities. We showed how the values
of the integrals relevant to determine the consistent models and
their spectra, can be obtained as integrals of divisors on the
corresponding toric variety. Unfortunately, only for the special 
$\Cplx^n/\Intr_n$ singularities explicit blowups are known; for more
complicated and phenomenologically more relevant orbifolds explicit
constructions remain a difficult task.

Luckily, toric geometry does not require that one has
explicitly constructed the metric of the non--compact Calabi--Yau
blowup of orbifold singularity: The geometrical orbifold action 
essentially uniquely determines the toric variety, that describes the
resolution of the orbifold singularity. The only caveat is, that the
resolution might not be topologically unique. The main advantage of
having the resolution of the orbifold compared to the orbifold itself,
is, that one is able to determine the structure inside the
singularity. This is encoded by the exceptional divisors, which were
needed to desingularize the toric variety. From the very definition of
these exceptional divisors it is clear, that they are in one--to--one
correspondence to the twisted sectors of orbifold string
theories. Motivated by this, we gave a self contained  introduction to
toric geometry for non--experts, emphasizing the methods relevant to
obtain heterotic models on toric orbifold resolutions. As it is
rather cumbersome to describe these procedures in general, we have
illustrated the toric geometrical tools by constructing heterotic
models on the resolutions of $\Cplx^2/\Intr_3\,$,  $\Cplx^3/\Intr_4$
and $\Cplx^3/\Intr_2\times\Intr_2'$ orbifolds. During our
investigations  the following issues came up:

We used the homogeneous coordinate approach to the construction of
toric varieties,  and the corresponding exclusion
set~\cite{Cox:1993fz}. We found however, that integrals of divisors,
that as hypersurfaces are excluded, can nevertheless give rise to
non--vanishing values. Already for the simple resolution of
$\Cplx^n/\Intr_n$ the intersection of all ordinary divisors, is part
of the exclusion set. However, both using linear equivalences, and 
integrating the corresponding background field strength on the
explicit blowup, we showed that such integrals are nonzero, but rather
fractional. Even though intersection theory of non--compact divisors
might be 
ill--defined\footnote{As D.\ Cox pointed out to us, the intersection of
non--compact divisors is problematic because the corresponding Chow
group is trivial.}, the integrals of the first Chern classes of the
line bundles associated to the divisors do give unambiguous
results in the cases considered. 
The reason is that the integrands are uniquely defined upto
exact terms, which means that the integrals over the non--compact
resolution are defined upto boundary terms. For applications
to blowups of compact orbifolds, one needs to glue various
non--compact resolutions together. The boundary contributions are then
canceled among themselves automatically, and the result is uniquely
defined. Hence, an alternative way to deal with this complication is
to consider the intersection theory of resolutions of compact 
orbifolds~\cite{Lust:2006zh,Reffert:2007im}.)

After these mathematical issues we turned to the applications in
heterotic model building. There are many consistency conditions, which
can be enforced on  heterotic models on the resolution of an
orbifold. First of all, there are the minimal requirements to
construct a sensible model on the resolution of the orbifold: The
$\U{1}$ gauge bundles have to be integral on all compact curves, both
in three dimensional complex resolutions and all compact curves of the two
dimensional subresolutions. In addition, the integrals of the Bianchi
identity over all compact exceptional divisors (compact four
dimensional real cycles) of the resolution have to vanish as well. To
be able to compute the spectrum of the model on the  
resolution, one needs to ensure, that the Bianchi identity integrated over
all non--compact $4$--cycles, and all subresolutions, i.e.\ the Bianchi
identity in six dimensions, vanish. Surprisingly, satisfying all these
conditions on the resolution of the orbifold seems to guarantee, that
in the blow down limit the model can be directly interpreted as a
heterotic orbifold. A direct identification of the orbifold gauge
shift vector with the $\U{1}$ gauge background can be obtained by
computing integrals over non--compact curves. By Stoke's theorem we
can turn it into a contour integral at infinity, which can be
identified with the same integral of the orbifold model.

For each of the resolution models we have computed the spectra. To
this end we used the conventional index theorem dropping possible
boundary contributions. This can be justified by imagining resolutions of
compact orbifolds: the boundary contributions from the 
local resolutions of the various fixed points precisely cancel in the
gluing procedure. In any event we have confirmed, that we are able to
reproduce the complete spectra of the heterotic orbifold models upto
vector--like matter. All in all we have obtained a detailed dictionary
of how to translate between orbifold and blowup model properties.

As explained above, not all requirements are  necessary, hence one may
wonder what happens if some of them are not fulfilled. In particular,
we could have non--vanishing Bianchi identities, when
integrated on non--compact $4$--cycles. This is very natural when one
thinks of obtaining blowup models of compact orbifolds: Then one only
has compact $4$--cycles; on each of them the integrated
Bianchi needs to vanish. From a local perspective this means that
there is $H$--flux exchanged between the resolutions of the
various fixed points. Using the results of~\cite{Lust:2006zh} one
should be able to analyze such situations globally. However, one
knows from orbifold field and string theory, that the spectra can be
determined locally at each of the fixed points (even in the presence
of Wilson lines). However, the standard index theorem, used in the
work, to compute the chiral spectrum fails, because it does not take
local $H$--fluxes into account. Using a modified index theorem, that
is valid in the presence of such fluxes, one may hope to be able to
compute the local spectra at any of the resolution models, that only
satisfy the necessary vanishing Bianchi conditions.

Another natural extension of our work, is to determine the blowup
models of the $T^6/\Intr_{6-II}$ orbifold. As was emphasized
in~\cite{Lebedev:2006tr,Lebedev:2006kn} such orbifolds with Wilson
lines seem to be able to give a relatively large class of MSSM--like
models. It would therefore be very interesting to study these models
in blowup. The $T^6/\Intr_{6-II}$ orbifold contains various orbifold
singularities, that are of the types $\Cplx^2/\Intr_2\,$,
$\Cplx^2/\Intr_3$ and $\Cplx^3/\Intr_{6-II}\,$. The construction of
resolution models for the first two singularities have been discussed
in this paper; for the first one we have constructed an explicit
blowup in~\cite{Nibbelink:2007rd}. The final singularity type can be
investigated using the methods explained here. In fact, there are five
topologically inequivalent resolutions and any resolution involves
four exceptional divisors. Therefore, each inequivalent resolution is
characterized by $20$ integrals number of the exceptional
divisors.  As the full analysis will therefore be rather involved, we 
postpone it to a future publication.

\subsection*{Acknowledgments}

We would like to thank O.\ Lebedev, H.P.\ Nilles and S.\ Raby for
stimulating discussions that encouraged us to pursue the investigation
of general orbifold resolution. We are grateful to D.\ Cox for
enlightening correspondence. And we would like to thank F.\ Pl\"oger
for reading the manuscript and giving useful suggestions.


\newpage 
\spectraCthreeZfour
\spectraCthreeZfourbis

\bibliographystyle{paper}
{\small
\bibliography{paper}
}
\end{document}